\documentclass[12pt]{article}
\usepackage{amsmath}
\usepackage{amsfonts}
\usepackage{amssymb}
\usepackage{dsfont}
\usepackage[latin1]{inputenc}
\usepackage[all]{xy}
\usepackage{graphicx}
\usepackage{latexsym}
\usepackage{color}
\usepackage{epsfig}
\usepackage{graphicx}
\usepackage{caption}
\usepackage{subcaption}
\usepackage{here}
\usepackage{float}
\usepackage{array}
\usepackage{tikz}
\usepackage{tikz-cd}
\usetikzlibrary{3d,arrows,calc,chains,matrix,positioning,scopes}

\def\l[{\left[}
\def\r]{\right]}
\newcommand{\slfrac}[2]{\left.#1\middle/#2\right.}

\newcommand{\tch}{{$\check{\mbox{C}}$}ech}

\def\ba{\begin{array}}
\def\ea{\end{array}}
\def\beq{\begin{equation}}
\def\eeq{\end{equation}}
\def\bea{\begin{eqnarray}}
\def\eea{\end{eqnarray}}

\newcommand{\Hom}{\mathop{}\mathopen{}{\rm Hom}\!}
\newcommand{\End}{\mathop{}\mathopen{}{\rm End}\!}
\newcommand{\Cupp}{{\rm{\smallsmile}}}
\newcommand{\Ext}{\mathop{}\mathopen{}{\rm Ext}}

\newcommand{\egzz}{\, \stackrel{\mathbb{Z}}{{=}}\,}

\begin{document}

\pagestyle{empty}
\setcounter{page}{0}
\hspace{-1cm}

\begin{center}
{\Large {\bf Abelian BF theory and Turaev-Viro invariant}}%
\\[1.5cm]

{\large P. Mathieu and F. Thuillier}

\end{center}

\vskip 0.7 truecm

{\it LAPTH, Universitée de Savoie, CNRS, 9, Chemin de Bellevue, BP 110, F-74941
Annecy-le-Vieux cedex, France.}

\vspace{3cm}

\centerline{{\bf Abstract}}

\noindent The $U(1)$ BF Quantum Field Theory is revisited in the light of Deligne-Beilinson Cohomology. We show how the $U(1)$ Chern-Simons partition function is related to the BF one and how the latter on its turn coincides with an abelian Turaev-Viro invariant. Significant differences compared to the non-abelian case are highlighted.

\vspace{2cm}
\noindent \textit{For Raymond Stora, in memoriam. The construction continues, but the edifice will never be the same shape.}
\indent

\vfill
\rightline{LAPTH-049/15}
\newpage
\pagestyle{plain} \renewcommand{\thefootnote}{\arabic{footnote}}

\section{Introduction}

\noindent Deligne-Beilinson Cohomology \cite{De,Be} takes its roots in Algebraic Geometry, and more specifically in the theory of Regulators and L-Functions as well as in K-Theory \cite{Ka}. It is also effective in the study of flat vector bundles \cite{EV,Ja} or in the classification of abelian Gerbes with connections \cite{MP}. Alternative descriptions are provided by Cheeger-Simons Differential Characters \cite{CS,Ko,Br}, Hopkins-Singer Differential Cohomology \cite{HS} and Harvey-Lawson Sparks \cite{HLZ}, all these notions being themselves equivalent in some sense \cite{SS}. These mathematical objects can be seen as refinement of the usual Characteristic Classes appearing in the attempt to classify vector bundles. For instance, the Deligne-Beilinson cohomology group $H_D^{1}(M,\mathbb{Z})$ of a closed manifold $M$ can be seen as the set of equivalence classes of $U(1)$-bundles with connection (i.e. gauge potential) over $M$. It is precisely this identification that allows to revisit and reinterpret the Ehrenberg-Siday-Aharonov-Bohm Effect \cite{ES,AB} thus providing the first clue of a possible use of Deligne-Beilinson Cohomology in the context of Quantum Physics. This is confirmed in the so-called Geometric Quantization \cite{Wo}.

The occurrence of the Chern-Simons Lagrangian -- also known as the Hopf or Whitehead Lagrangian in the abelian case -- in Quantum Physics is also a quite old story \cite{Sc,Sch,DJT,WZ,Ha,Po,Wi0,Wil} and similarly for the "brother" BF Lagrangian \cite{H,BT,BBRT,MS,CCFMRT}. For instance the Chern-Simons Lagrangian has been used in (2+1)-dimensional topological gravity \cite{JP} and has provided arguments for developments in Loop Quantum Gravity \cite{Wi2}, a theory in which the BF Lagrangian appears to play also an important role \cite{P}. The $U(1)$ Chern-Simons and BF Lagrangians have also proven fruitful in Condensed Matter physics and more specifically in the description of Quantum Hall Effect \cite{GMD}, Superconductors \cite{HOS} and Topological Insulators \cite{CM}. Although the relationship of these Lagrangians with Differential Characters was known to mathematicians since the mid-70s \cite{CS,Ko}, this aspect of things was most of the time ignored by physicists mainely because Quantum Physics usually deals with $\mathbb{R}^3$ and not with generic 3-dimensional manifolds.

In 1989 E. Witten \cite{Wi} made a breakthrough in the understanding of the non-abelian Chern-Simons (CS) Quantum Field Theory by showing how it is related to Knot Theory \cite{Ro} in dimension $3$. Up to some normalization Witten's article points out how the expectation values of Wilson loops in the $SU(2)$ Chern-Simons theory identify themselves with the Jones polynomials \cite{Jo}. Shortly after \cite{GMM} this was perturbatively confirmed, that is to say by using Quantum Field Theory technics \cite{Gu}. It was also observed at that time that the Chern-Simons Lagrangian can be seen as a Deligne-Beilinson cohomology class of degree $3$ and similarly that the Wess-Zumino-Witten term is a Deligne-Beilinson cohomology class of degree $2$ on the associated Lie group \cite{DW,CJMSW}. This Wess-Zumino-Witten term arises when checking the gauge invariance of the Chern-Simons theory at the quantum level which involves quantization of the coupling constant of this theory. However the role of Deligne-Belinson cohomology at the level of quantum fields is not evident in the non-abelian context. Nevertheless some recent results suggest that even in the non-abelian framework Deligne-Belinson cohomology could help to understand the Chern-Simons theory \cite{FSS}. As for the BF Quantum Field Theory, its relation with knots and links was also stressed out in some articles \cite{CCM,CCFM}. In \cite{CCFM} it was shown at the level of functional integration how the partition function of the $SU(n)$ BF model with cosmological constant coincides with the absolute square of the $SU(n)$ Chern-Simons partition function for an appropriate choice of coupling constants. It is one of the aims of the present article to check whether this property holds true in the $U(1)$ case.

Although the generic role of Deligne-Beilinson Cohomology in Quantum Field Theory was stressed out in some articles \cite{Ga,BGST}, it is in the context of the $U(1)$ Chern-Simons Quantum Field Theory that this role has been specially relevant. It was shown that all the results usually obtained by surgery arguments can be recovered from a functional integration point of view, even in dimensions $4n+3$ \cite{GPT}, once Deligne-Beilinson Cohomology is introduced into the game \cite{GT1,T1,GT2,GT3}. For instance the use of the Deligne-Beilinson cohomology group $H_D^{1}(M,\mathbb{Z})$ as configuration space implies quantization of the coupling constant and charges of the theory albeit no Wess-Zumino term occurs in the abelian context. It was also proven that for any oriented closed $3$-manifold $M$ the partition function of the $U(1)$ Chern-Simons coincides, up to a normalization which is universal in its form, with a Reshetikhin-Turaev (RT) invariant of $M$ \cite{GT3}. Eventually some kind of "algebraic surgery" ermerges thus allowing to express all the results obtained in $M$ as ones in $S^3$ \cite{T2}.

Reshetikhin-Turaev invariants, originally based on Hopf algebra and Quantum groups \cite{RT}, were introduced with the intention to get a better understanding of Jones polynomials and in the hope to obtain new invariant polynomials. These invariants are now well understood in the more abstract language of modular category \cite{T}. Another set of topological invariants for closed 3D manifolds, named Turaev-Viro (TV) invariants, was introduced by using $6$-j symbols and Quantum Groups \cite{TV}. As for RT ones, the construction of TV invariants can be rephrase in the language of categories, the spherical ones \cite{GK,BW,BK}. These two sets of invariant are actually related since for a given modular category the absolute square of the RT invariant is equal to the TV invariant \cite{T}. It has to be pointed out that in the $SU(2)$ case the TV invariant can be seen as a regularization of the Ponzano-Regge formula \cite{PR}, itself being presented as a discretization of the $SU(2)$ BF theory \cite{Ba}. It is another aim of this article to check whether the $SU(2)$ BF theory is also associated with a TV invariant. Deligne-Beilinson cohomology is used in order to explicitly compute the partition functions of the $U(1)$ CS and BF Quantum Field Theory and then to check if the absolute square of the former equals the latter.

Strictly speaking it is smooth Deligne cohomology which is involved in the article. Nonetheless we keep the terminology Deligne-Beilinson cohomology for the sake of continuity with the choice originally made in \cite{BGST}. Moreover Deligne-Belinson cocycles actually depend on two integers: one defining the length of the Deligne-Belinson complex (which is where the de Rham complex is truncated) and the second being the degree of the cohomology under consideration. We only consider the case where these two integers are equal. Note also that most results concerning CS have already been obtained (see \cite{GT3,T2} for instance).

\vspace{2mm}
\noindent In Subection 2.1 and 2.2 some basic facts about DB cohomology are recalled and applied in the context of $U(1)$ CS and BF theories which yields:
\vspace{2mm}

\noindent {\bf Lemma 1.} {\it Once the configuration space of the $U(1)$ BF theory is chosen as (an appropriate subset of) the Deligne-Beilinson cohomology groups product $H_D^1(M,\mathbb{Z})^\ast \times H_D^1(M,\mathbb{Z})^\ast$, the Lagrangian is given as a Deligne-Beilinson product, thus implying quantization of the coupling constant ($k \in \mathbb{Z}$).}
\vspace{2mm}

\noindent Subsection  2.3 is dedicated to the introduction and study of partition functions thus leading to:
\vspace{2mm}

\noindent {\bf Lemma 2.} {\it For a given coupling constant $k$:

\noindent 1) The $U(1)$ BF theory is $k$-periodic whereas the $U(1)$ Chern-Simons theory is $2k$-periodic.

\noindent 2) Like in the $U(1)$ Chern-Simons theory only the torsion sector of $H_1(M)$ contributes to the $U(1)$ BF partition function.

\noindent 3) Unlike the non-abelian case, the partition function of the $U(1)$ BF theory is not always the square norm of the CS partition function. More precisely, consider the standard decomposition $T_1(M) = \bigoplus\limits_{j=1}^{d} \mathbb{Z}_{p_j}$ of the torsion part of $H_1(M)$ (with $p_j | p_{j+1}$), set $p'_j = p_j/gcd(k,p_j)$, denote by $\beta$ (resp. $\gamma$) the number of $j$ such that $p'_j = 2(2l_j +1)$ (resp. $p'_j = 4 l_j$), then:
\bea
\left| Z_{CS_{k}}(M) \right|^2  =  \delta_{\beta, 0} \, 2^\gamma \, Z_{BF_{k}}(M) = {2^{-\beta} \delta_{\beta, 0} \, Z_{BF_{2k}}(M)}  \, .
\eea
}
\vspace{2mm}

\noindent In Section 3 the relation between the CS and BF partitions functions with respectively the RT and TV invariants is investigated thus yielding the last series of results gathered into:
\vspace{2mm}

\noindent {\bf Lemma 3.}

\noindent {\it 1) The absolute square of the $U(1)$ CS partition function is related to the absolute square of the Reshetikhin-Turaev invariant according to:
\bea
|\tau_{4k}(M)|^2 = \frac{(2k)^{b_1}}{p_1 \ldots p_d} \, |Z_{CS_{k}}(M)|^2 \, .
\eea
In particular, the RT invariant $\tau_{4k}(M)$ built from the non modular category $\mathbb{Z}_{4k}$ admits a reduced expression in which charges take their values in $\mathbb{Z}_{2k}$ instead of $\mathbb{Z}_{4k}$.

\noindent 2) There is a natural abelian Turaev-Viro invariant, $\Upsilon_k(M)$, the construction of which relies on the spherical category $\mathbb{Z}_k$ and which is related to the $U(1)$ BF partition function $Z_{BF_{k}}(M)$ according to:
\bea
\label{lemmaTvvsBF}
\Upsilon_k(M) = |H^1(M,\mathbb{Z}_{k})| = \frac{k^{b_1}}{p_1 \ldots p_d} Z_{BF_{k}}   \, .
\eea

\noindent Result 3) of Lemma 2 implies that:
\bea
\Upsilon_{2k}(M) \neq |\tau_{4k}(M)|^2 \, .
\eea

\noindent 3) When considering the modular category $\mathbb{Z}_{2k+1}$ we recover that:
\bea
|\tau_{2k+1}(M)|^2 = \Upsilon_{2k+1}(M) = \frac{(2k+1)^{b_1}}{p_1 \ldots p_d} Z_{BF_{2k+1}}  \, .
\eea
However there is no $U(1)$ CS theory associated with this RT invariant $\tau_{2k+1}(M)$.

\noindent 4) The semi-simple category $\mathbb{Z}_{2(2k+1)}$ gives a mix of results 2) and 3):
\bea
\Upsilon_{2(2k+1)}(M) = \frac{(2(2k+1))^{b_1}}{p_1 \ldots p_d} Z_{BF_{k}}  \, ,
\eea
and there are neither RT invariant nor $U(1)$ CS theory associated with $\mathbb{Z}_{2(2k+1)}$.}

\vspace{5mm}

\noindent All along this article CS and DB will stand for Chern-Simons and Deligne-Beilinson respectively, and $\egzz$ will mean equality modulo $\mathbb{Z}$. Furthermore when refering to "non-abelian" we mean that a simply connected group like $SU(2)$ (or $SU(n)$) is involved.

\section{The $U(1)$ BF Quantum Field Theory in the Deligne-Beilinson framework}

\noindent As mentioned in the introduction Deligne-Beilinson cohomology provides a new angle under which to see some quantum theoretical problems. If historically the Ehrenberg-Siday-Aharonov-Bohm effect can be considered as the first example of this new enlightenment, Chern-Simons theory is one of those where DB cohomology proved itself particularly successful. Among other topological models the $U(1)$ BF one is, due to the nature of its Lagrangian, the closest to the CS theory. It is the aim of this section to investigate this relation into details. In the standard $U(1)$ BF Quantum Field Theory on $\mathbb{R}^3$ the Lagrangian is assumed to be $A \wedge dB$ for two $U(1)$ gauge fields on $\mathbb{R}^3$. Whereas these gauge fields are identified with $1$-forms on $\mathbb{R}^3$, this identification is not possible on a generic closed manifold. So our first task will be to determine the most appropriate configuration space on which to consider the $U(1)$ BF Lagrangian. As a consequence we will recover that on a closed $3$-manifold the gauge group identifies itself with closed $1$-forms with integral periods instead of the group of exact $1$-forms as it happens in $\mathbb{R}^3$.

\subsection{Deligne-Beilinson cohomology as configuration space}

\noindent The simplest example of Deligne-Beilinson cocycles is provided by $U(1)$-connections on $U(1)$-principal bundles over an oriented closed smooth manifold $M$. The corresponding cohomology group is then a way to describe the set of equivalent classes of $U(1)$-principal bundles with connection over $M$.

\medskip

\noindent (1) {\it Connections and Deligne-Beilinson cocycles.} A cover $\mathfrak{U}$ of $M$ is good if any non empty intersection of open sets of $\mathfrak{U}$ is diffeomorphic to $\mathbb{R}^n$, with $n = dim M$. Given a good cover $\mathfrak{U} = (U_{\alpha})_{\alpha \in I}$ of $M$, a $U(1)$-connection is a collection of triples $\mathbf{A} = (A_\alpha , \Lambda_{\alpha \beta} , n_{\alpha \beta \gamma})$ such that the $1$-forms $A_\alpha$, the functions $\Lambda_{\alpha \beta}$ and the integers $n_{\alpha \beta \gamma}$ fulfill:
\bea
\label{DBcocyle}
\left\{ \begin{gathered}
  (\delta_0 A)_{\alpha \beta} := A_\beta - A_\alpha = d_0 \Lambda_{\alpha \beta} \hfill \\
  (\delta_1 \Lambda)_{\alpha \beta \gamma} := \Lambda_{\beta \gamma} - \Lambda_{\alpha \gamma} + \Lambda_{\alpha \beta} = d_{-1} n_{\alpha \beta \gamma} \hfill \\
  (\delta_2 n)_{\alpha \beta \gamma \rho} := n_{\beta \gamma \rho} - n_{\alpha \gamma \rho} + n_{\alpha \beta \rho} - n_{\alpha \beta \gamma} = 0 \hfill \\
\end{gathered}  \right. \, ,
\eea
respectively in all $U_{\alpha \beta}$, $U_{\alpha \beta \gamma}$ and $U_{\alpha \beta \gamma \rho}$. As usual $U_{\alpha_0 \cdots \alpha_n}$ denotes the (non empty) intersection $U_{\alpha_0} \cap \cdots \cap U_{\alpha_n}$ and $d_{-1}$ is the canonical injection of numbers into (constant) functions, $d_{-1} : \mathbb{R} \rightarrow \Omega^0(M)$. By setting $g_{\alpha \beta} := e^{2 i \pi \Lambda_{\alpha \beta}}$ we can rebuild a $U(1)$-principal (coordinate \cite{St}) bundle $P$ over $M$ with transition functions $g_{\alpha \beta}$ and where the $1$-forms $A_{\alpha}$ are the pull-back of a $U(1)$-connection $\mathcal{A}$ on $P$ under some local sections $A_{\alpha} = s_{\alpha}^* \mathcal{A}$. The collection $\mathbf{A}$ is also called a \tch-de Rham representative of a DB $1$-cocycle. Two DB cocycles $\mathbf{A}$ and $\mathbf{\tilde{A}}$ are said to be equivalent (or cohomologous) if:
\bea
\label{DBboundary}
\left\{ \begin{gathered}
\tilde{A}_{\alpha} - A_{\alpha} =  d_0 q_{\alpha} \hfill \\
\tilde{ \Lambda}_{\alpha \beta} - \Lambda_{\alpha \beta} =  (\delta_0 q)_{\alpha \beta} + d_{-1} m_{\alpha \beta} \hfill \\
\tilde{n}_{\alpha \beta \gamma} - n_{\alpha \beta \gamma} = (\delta_1 m)_{\alpha \beta \gamma} \hfill \\
\end{gathered}  \right. \, .
\eea
The equivalence class of a DB cocycle $\mathbf{A}$ is denoted by $\bar{{\bold A}}$. It can be shown that this construction is of cohomology type \cite{EV,BGST}.

\medskip

\noindent (2) {\it Deligne-Beilinson cohomology groups.} Extending the previous construction to forms of higher degree leads to $U(1)$ Gerbes with connections over $M$ which generalize $U(1)$-principal bundles with connections. The corresponding DB cohomology groups appear as the sets of equivalence classes of $U(1)$ Gerbes with connections \cite{Br,MP}.

In the case of an oriented closed 3-manifold $M$ the DB cohomology space $H_D^1 \left({M,{\mathbb Z}} \right)$ is canonically embedded into the following exact sequence \cite{CS,Br,HLZ,BGST}:
\begin{equation}
\label{1}
0 \longrightarrow {\Omega^1( M) \over {\Omega_{\mathbb{Z}}^1(M)}} \longrightarrow H_D^1(M,\mathbb{Z})
\longrightarrow H^{2}(M,\mathbb{Z}) \longrightarrow 0 \, ,
\end{equation}
where $\Omega ^1\left( M \right)$ is the space of smooth 1-forms on $M$,
$\Omega _{\mathbb Z}^1 \left( M \right)$ the space of smooth closed 1-forms with
integral periods on $M$ and $H^{2}\left( {M,{\mathbb Z}} \right)$ is the
second ($\check{\textrm{C}}$ech) cohomology group of $M$. Note that $\Omega _{\mathbb Z}^1 \left( M \right)$ is nothing but the global gauge group of $U(1)$-connections on $M$. There is another exact sequence into which $H_D^1 \left({M,{\mathbb Z}} \right)$ can be embedded (\cite{CS,HLZ}):
\begin{equation}
\label{2}
0 \longrightarrow H^{1}(M,\mathbb{R}/\mathbb{Z}) \longrightarrow H_D^1(M,\mathbb{Z}) \longrightarrow \Omega_{\mathbb{Z}}^2(M) \longrightarrow 0 \, ,
\end{equation}
where $H^{1}(M,\mathbb{R}/\mathbb{Z})$ is the first $\mathbb{R
}/\mathbb{Z}$-valued ($\check{\textrm{C}}$ech) cohomology group of $M$ and $\Omega _{\mathbb Z}^2(M)$ the space of smooth closed 2-forms with integral periods on $M$.

\begin{figure}
  \centering
  \includegraphics[scale=0.5]{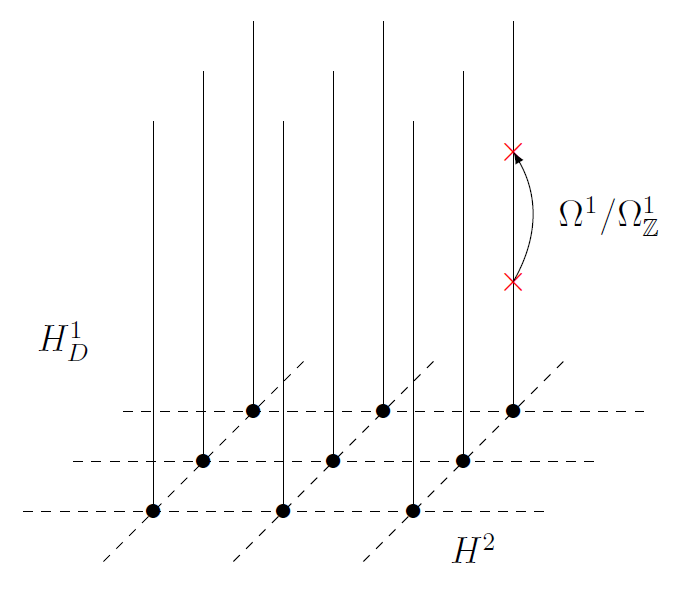}
  \caption{Representation of the first exact sequence.}\label{fig1}
\end{figure}

\begin{figure}
  \centering
  \includegraphics[scale=0.4]{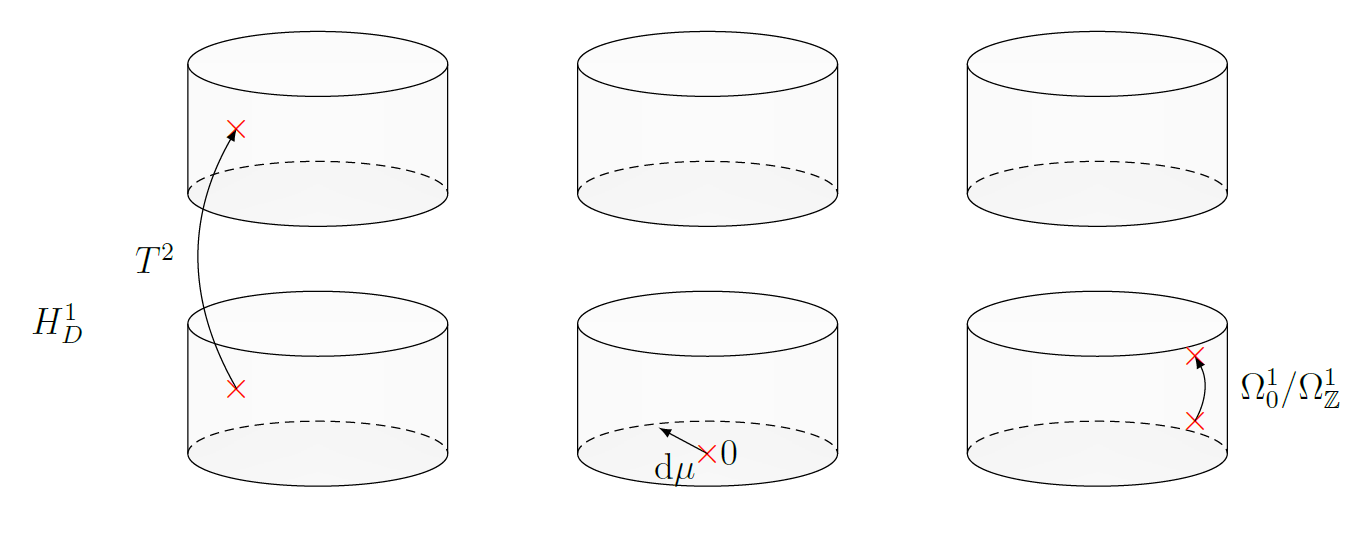}
  \caption{Representation of the second exact sequence.}\label{fig2}
\end{figure}

Each one of these two exact sequences has its own interest to describe $H_D^1(M,\mathbb{Z})$, but both give this space the structure of an affine bundle, with (discrete) base $H^{2}(M,\mathbb{Z})$  and translation group $\Omega^1(M) / \Omega_{\mathbb Z}^1(M)$  for the former sequence (Figure \ref{fig1}), and with base $\Omega_\mathbb{Z}^2(M)$ and translation group $H^{1}(M,\mathbb{R}/\mathbb{Z})$ from the latter one (Figure \ref{fig2}). As we will see sequence (\ref{1}) turns out to be the best suited for the BF theory (as it was for the Chern-Simons theory). Nevertheless sequence (\ref{2}) is the one that physicists like the most because curvatures, i.e. closed 2-forms with integral periods, are clearly identified in this sequence. There is a simple drawing that helps to convince oneself that the two exact sequences are equivalent.

\begin{figure}
  \centering
  \includegraphics[scale=0.43]{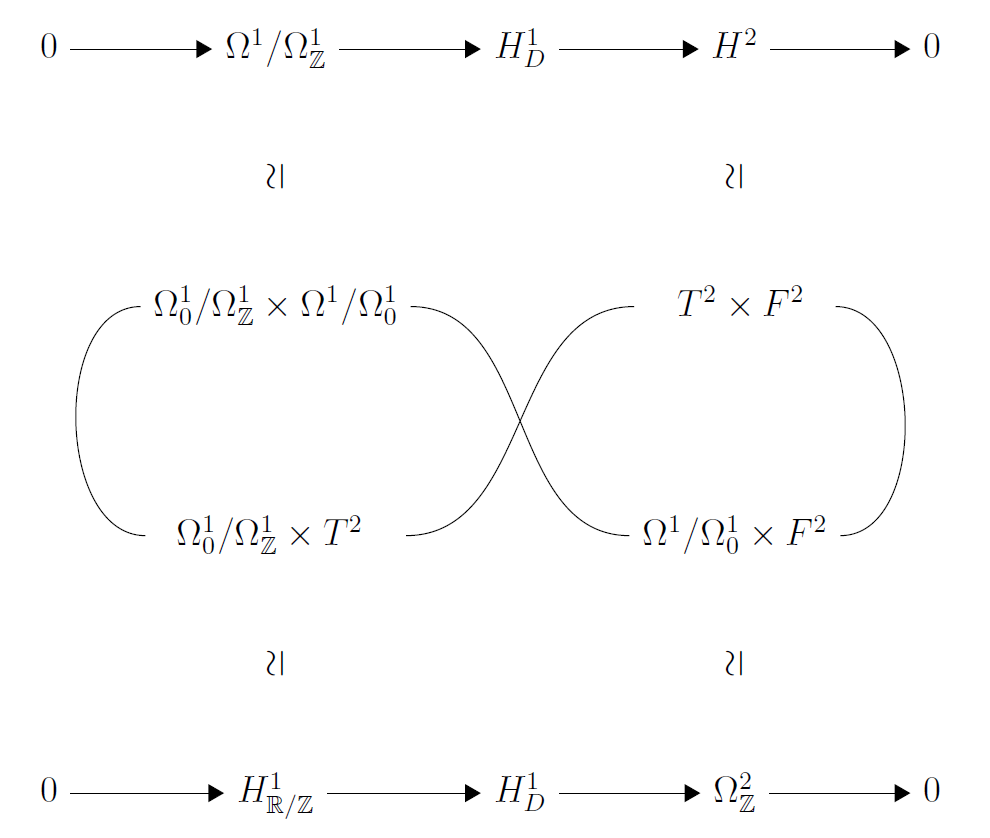}
  \caption{Information in the two exact sequences.}\label{fig3}
\end{figure}

We have $H^{1}(M,\mathbb{R}/\mathbb{Z}) \simeq T^2(M) \times (\mathbb{R}/\mathbb{Z})^{b_1}$ where $b_1$ is the first Betti number, that is to say the dimension of $H_1(M,\mathbb{R}) \simeq H^2(M,\mathbb{R})$, and $T^2(M)$ is the torsion sector of $H^2(M,\mathbb{Z})$ . On the other hand we can check that $\Omega _{\mathbb Z}^2 \left( M \right) \simeq F^2(M) \times ({\Omega ^1(M)} / {\Omega_{0}^1(M)})$ where $F^2(M)$ denotes the free sector of $H^2(M,\mathbb{Z})$ and ${\Omega_{0}^1(M)}$ the space of closed $1$-forms on $M$ (Figure \ref{fig3}).

\medskip

\noindent (3) {\it Pontrjagin duality and holonomy.} At this stage it would be natural to consider $H_D^1(M,\mathbb{Z})$ as our configuration space. However it is a well known fact of Quantum Field Theory that quantum fields turn out to be currents (i.e. forms with distribution coefficients) rather than smooth forms. The space of $p$-currents is usually defined as the topological dual of $\Omega^{n-p}(M)$, the space of $(n-p)$-forms on $M$. However due to the background presence of $U(1)$ (which appears here as $\mathbb{R}/\mathbb{Z}$) it would be logical to consider the Pontrjagin dual of $H^{1}(M,\mathbb{R}/\mathbb{Z})$: $H_D^1(M,\mathbb{Z})^{\ast} \equiv Hom\left( {H_D^1 \left( {M,{\mathbb Z}} \right),{{\mathbb R}/{\mathbb Z}}} \right)$ as our configuration space. It turns out that $H_D^1(M,\mathbb{Z})^{\ast}$ contains $H_D^1(M,\mathbb{Z})$ -- in the same sense that the space of $1$-currents on $M$ contains $\Omega^{1}(M)$ -- but also the space $Z_1(M)$ of $1$-cycles on $M$. The inclusion $Z_1(M) \subset H_D^1(M,\mathbb{Z})^{\ast}$ is straightforwardly realized by the pairing:
\begin{equation}
\label{11}
\oint : H_D^1 \left( {M,{\mathbb Z}} \right) \times Z_1 \left( {M}
\right)\longrightarrow {{\mathbb R}/{\mathbb Z}} \, ,
\end{equation}
A precise definition of integration over $1$-cycles is provided by the \tch-de Rham writting for representatives of DB classes \cite{BGST,T2}. For the DB class $\bar{{\bold A}}$ of $\mathbf{A} = (A_\alpha , \Lambda_{\alpha \beta} , n_{\alpha \beta \gamma})$ let $z$ be a $1$-cycle on $M$ such that with respect to the good cover $\mathfrak{U}$ this cycle gives rise to the collection $(z_\alpha , x_{\alpha \beta})$ such that $\sum_\alpha z_\alpha = z$ and $\sum_\beta  x_{\alpha \beta} = b z_\alpha$, where $b$ denotes the boundary operators on chains. Although not all $1$-cycles admit such a decomposition, for a given set of $1$-cycles it is always possible to find a good cover for which all these $1$-cycles admit such a decomposition \cite{We}. Then:
\bea
\label{2.43}
\oint_z \bar{\mathbf{A}}  \egzz \sum_\alpha \int_{z_\alpha} A_\alpha - \sum_{\alpha , \beta} \int_{x_{\alpha \beta}} \Lambda_{\alpha \beta} \, ,
\eea
It can be checked that modulo integers expression (\ref{2.43}) is independent of the representative of $\bar{\mathbf{A}}$ as well as of the collection $(z_\alpha , x_{\alpha \beta})$ representing the $1$-cycle $z$.

On the other hand, by taking the $\Hom \;$ of the exact sequences (\ref{1})-(\ref{2}) we obtain the dual sequences \cite{HLZ}:
\begin{equation}
\label{11b}
0 \longrightarrow H^{1}(M,\mathbb{R}/\mathbb{Z}) \longrightarrow H_D^1(M,\mathbb{Z})^{\ast} \longrightarrow \Hom \left(  {\Omega^1(M) \over \Omega_{\mathbb{Z}}^1(M)} \right) \longrightarrow 0 \, ,
\end{equation}
and
\begin{equation}
\label{12}
0\buildrel \over \longrightarrow \Hom \left( \Omega _{\mathbb Z}^2 \left( M \right),{{\mathbb R}/{\mathbb Z}} \right)\buildrel
 \over \longrightarrow H_D^1(M,\mathbb{Z})^{\ast} \longrightarrow H^{2}(M,\mathbb{Z}) \longrightarrow 0 \, ,
\end{equation}
These sequences are very similar to the original ones: the second one presents $H_D^1(M,\mathbb{Z})^{\ast}$ as a discrete bundle with base $H^{2}(M,\mathbb{Z})$ just like $H_D^1(M,\mathbb{Z})$. Note that the action of $\Hom$ exchange the role of the two sequences.
Exterior product and integration over $M$ provide canonical injections:
\bea
\label{12bis}
\Omega_{\mathbb Z}^2(M) &\hookrightarrow& \left( {{\Omega ^1(M)} \over {\Omega_{\mathbb{Z}}^1(M)}} \right)^* := \Hom\left(  {{\Omega ^1(M)} \over {\Omega_{\mathbb{Z}}^1(M)}},{{\mathbb R}/{\mathbb Z}} \right) \, ,
\\ \nonumber
{{\Omega^1(M)} \over {\Omega_{\mathbb Z}^1(M)}} &\hookrightarrow& \left( \Omega_{\mathbb Z}^2(M) \right)^* := \Hom\left( \Omega _{\mathbb Z}^2(M),{{\mathbb R}/{\mathbb Z}} \right)
\, ,
\eea
which allow to show that:
\bea
\label{12quart}
Z_1(M) \oplus H_D^1(M,\mathbb{Z}) \hookrightarrow H_D^1(M,\mathbb{Z})^{\ast}
\, ,
\eea
as announced. This direct sum can be considered as the simplest extension of the smooth configuration space $H_D^1(M,\mathbb{Z})$. In the case of $U(1)$ Chern-Simons theory it turns out to be enough \cite{GT1,GT2,GT3,GPT}. Note that inclusion $Z_1(M) \subset H_D^1(M,\mathbb{Z})^{\ast}$ allows to see knots and links as elements of $H_D^1(M,\mathbb{Z})^{\ast}$ that is to say as Quantum Fields. In particular it was shown \cite{BGST} that we can associate to any $1$-cycle $z$ in $M$ a distributional DB class $\bar{\eta}_z$.
Another standard problem of QFT is the regularization of product of distributions. If we consider the simplest configuration space $Z_1(M) \oplus H_D^1(M,\mathbb{Z})$ we can already foresee that the only regularization will be those coming from "products" of cycles, the other fields being smooth. We will return to this in the next subsection about Lagrangians.

\medskip

\noindent (4) {\it Trivial, free and torsion origins.} Let's have a closer look at the structure of the configuration space as it appears from exact sequence (\ref{1}) (or alternatively (\ref{12})). We already mentioned that from this exact sequence point of view, $H_D^1(M,\mathbb{Z})$ is a discrete affine bundle over $H^2(M,\mathbb{Z})$. Hence once an origin on a fiber over $\vec{n} \in H^2(M,\mathbb{Z})$ has been chosen, all the elements on this fiber are reached through a translation by an element of ${\Omega^1(M) / {\Omega_{\mathbb{Z}}^1(M)}}$. On the fiber over $\vec{0}$ we can canonically chose as origin $\bar{\mathbf{0}}$, the class of the zero connection, that is to say the class of the DB cocycle $(0,0,0)$. With this canonical choice of {\bf trivial origin} the fiber over $\vec{0}$ can be identified with ${\Omega^1( M) / {\Omega_{\mathbb{Z}}^1(M)}}$ itself by setting: $\bar{\mathbf{A}} = \bar{\mathbf{0}} + \bar{\omega}$, where $\bar{\omega}$ denotes the class of $\omega \in \Omega^1(M)$ in the quotient ${\Omega^1(M) / {\Omega_{\mathbb{Z}}^1(M)}}$. With respect to the {\tch}-de Rham approach the collection $(\omega|_{U_{\alpha}} , 0 , 0)$ is a representative of $\bar{\omega}$ and hence any element of the fiber over $\vec{0}$ admit such a representative. The fiber over $\vec{0}$ will be called the {\bf trivial fiber}. Of course a trivial fiber also exists in $H_D^1(M,\mathbb{Z})^\ast$.

For the other fibers there are no such canonical choice of origins. As $Z_1(M) \subset H_D^1(M,\mathbb{Z})^\ast$ and since by Poincaré duality $H^2(M,\mathbb{Z}) \simeq H_1(M,\mathbb{Z})$ we can chose as origin on a fiber over $\vec{n}$ any $1$-cycle whose homology class is $\vec{n}$. Furthermore we have the $\mathbb{Z}$-modules isomorphism $H_1(M,\mathbb{Z}) \simeq \mathbb{Z}^{b_1} \oplus \mathbb{Z}_{p_1} \oplus \cdots \oplus \mathbb{Z}_{p_d}$, with $p_i | p_{i+1}$. The integer $b_1$ is the first Betti number of $M$ and it is the rank of $F_1(M)$, the free part of $H_1(M,\mathbb{Z})$, whereas the rest forms the torsion sector $T_1(M)$. Fibers over elements of $F_1(M) \simeq F^2(M)$ will be call {\bf free fibers} and those over elements of $T_1(M) \simeq T^2(M)$ will be called {\bf torsion fibers}. Any other fiber will be referred as a {\bf generic fiber}. Let $z_a$, $a = 1, \cdots ,b_1$, be a set of \emph{chosen} $1$-cycles on $M$ which generate $F_1(M)$. Then each $1$-cycle $\sum u^a z_a$ is taken as origin of the fiber over $(u^1, \cdots , u^{b_1}) \in F_1(M) \simeq \mathbb{Z}^{b_1}$. These origins will be referred as {\bf free origins} and will be denoted by $\bar{\mathbf{A}}_{\vec{u}}$. For $\tau \in \mathbb{Z}_{p_i}$, if the collection of integers $n_{\alpha \beta \gamma}$ is a {\tch} representative of the Poincaré dual of $\tau$ and if the collection of integers $m_{\alpha \beta}$ defines a {\tch} cochain $m$ such that $p_i.m = n$ then $(0 , d_{-1}(m_{\alpha \beta}/p_i), n_{\alpha \beta \gamma})$ is a representative of a DB class that belongs to the fiber over $\tau$  \cite{GT3,T2}. Although these origins are not canonical in general -- there may be two such classes on a given torsion fiber -- they will turn out to be very useful. They will be called {\bf torsion origins} and denoted by $\bar{\mathbf{A}}_{\vec{\kappa}}$ for $\vec{\kappa} \in T_1(M)$. Finally on a generic fiber over $\vec{u} + \vec{\kappa} \in H_1(M)$ we choose as origin the combination $\bar{\mathbf{A}}_{\vec{u}} + \bar{\mathbf{A}}_{\vec{\kappa}}$.

\vspace{2mm}

\noindent {\bf Examples:} 1) Since $H^2(S^3) = 0$ the space $H_D^1(S^3,\mathbb{Z})$ has only one fiber, the one over $\vec{0}$. Moreover, since $H^1(S^3) = 0$, $\Omega_{\mathbb{Z}}^1(M) = d \Omega_{\mathbb{Z}}^0(M)$. Hence any element of $H_D^1(S^3,\mathbb{Z})$ can be represented by a $1$-form which is unique up to an exact contribution. This is the closest case to what is usually considered in the BF Quantum Field Theory, as $S^3 = \mathbb{R}^3 \cup \{\infty\}$ (see next subsection).

2) Since $H^2(S^1 \times S^2) = \mathbb{Z} \simeq H_1(S^1 \times S^2)$ the space $H_D^1(S^3,\mathbb{Z})^\ast$ has only free fibers. We consider a $S^1$ generating $H_1(S^1 \times S^2)$ and take the cycle $n \cdot S^1$ as origin on the fiber over $n$, for any $n \in \mathbb{Z}$.

3) Since the first homology group of the lens spaces $L(p,q)$ is $\mathbb{Z}_p$ the configuration space on this manifold is made of torsion fibers only. Then we consider a generator $\tau$ of $\mathbb{Z}_p$ and the the DB class $\bar{\mathbf{A}}_{\vec{\tau}}$ of the canonical representative $(0 , d_{-1} (\zeta / p) , \tau)$ as torsion origin over $\tau$. On all the fibers we just pick up the DB classes $\bar{\mathbf{A}}_{n \cdot \vec{\tau}}$ of $(0 , n \cdot d_{-1} (\zeta / p) , n \cdot \tau)$ for $n = 0, \cdots , p-1$. We use the same letters for chains and cochains which is justified by Poincaré duality.

\vspace{2mm}

The other important DB space which is used is $H_D^3(M,\mathbb{Z})$. As $H^3(M,\mathbb{Z}) \simeq \mathbb{Z}$, the previous exact sequences give $H_D^3(M,\mathbb{Z}) \simeq ({\Omega^3(M)} / {\Omega_{\mathbb Z}^3(M)}) \simeq {\mathbb{R}/\mathbb{Z}} \simeq H^3(M,\mathbb{R}/\mathbb{Z})$. A simple way to realise this isomorphism is to fix a normalised volume form $\mu_0$ on $M$ such that for any $\theta \in {\mathbb R}/{\mathbb Z}$ the $3$-form $\theta \mu_0$  defines a unique DB element of $H_D^3 \left({M,{\mathbb Z}} \right)$ .  Note that  $H_D^3 \left( {M,{\mathbb Z}} \right)^{\ast} \cong  H_D^3 \left( {M,{\mathbb Z}} \right)^{\ast} \simeq {{\mathbb R}/{\mathbb Z}}$.

Before considering the construction of the Lagrangian and action defining the $U(1)$ BF theory it must be emphasized that DB cohomology spaces are $\mathbb Z$-modules. Hence we can only consider integral combinations of DB classes and a DB class can only be divided by $\pm 1$.

\subsection{Lagrangians, functional measures and periodicities}

\noindent The construction of the $U(1)$ BF and CS Lagrangians relies on the graded pairing between DB cohomology spaces.

\medskip

\noindent (1) {\it Deligne-Beilinson product.} For any smooth closed oriented 3-manifolds this pairing gives rise to the commutative product:
\begin{equation}
\label{4}
\star : H_D^1 \left( {M,{\mathbb Z}} \right) \times H_D^1 \left( {M,{\mathbb Z}}
\right)\longrightarrow H_D^{3} \left( {M,{\mathbb Z}}
\right)\simeq {{\mathbb R}/{\mathbb Z}} \, .
\end{equation}
Let $\mathbf{A} = (A_\alpha , \Lambda_{\alpha \beta} , m_{\alpha \beta \gamma})$ and $\mathbf{B} = (B_\alpha , \Theta_{\alpha \beta} , n_{\alpha \beta \gamma})$ be two $U(1)$ connections on $M$, viewed as \tch-de Rham representatives of the DB classes $\bar{\mathbf{A}}$ and $\bar{\mathbf{B}}$ respectively. The collection:
\bea
\label{prodrep}
(A_\alpha \wedge d B_\alpha \; , \; \Lambda_{\alpha \beta} \wedge d B_\beta \; , \; m_{\alpha \beta \gamma} . B_\gamma \; , \; m_{\alpha \beta \gamma} . \Theta_{\gamma \rho} \; , \; m_{\alpha \beta \gamma} .n_{\gamma \rho \epsilon})
\eea
is then a representative of the DB class $\bar{\mathbf{A}} \star \bar{\mathbf{B}}$. This expression provides a realisation of the DB product at the level of \tch-de Rham representatives.
\vspace{2mm}

\noindent {\bf Examples:} 1) Consider the classes $\bar{\omega}$ and $\bar{\chi}$ of the DB cocycles $(\omega|_{U_{\alpha}} , 0 , 0)$ and $(\chi|_{U_{\alpha}} , 0 , 0)$ where $\omega$ and $\chi$ are two $1$-forms on $M$. Then expression (\ref{prodrep}) reduces to $(\omega|_{U_{\alpha}} \wedge d \chi|_{U_{\alpha}} \; , \; 0 \; , \; 0 \; , \; 0 \; , \; 0) = ((\omega \wedge d \chi)|_{U_{\alpha}} \; , \; 0 \; , \; 0 \; , \; 0 \; , \; 0)$. This defines the DB product on $({\Omega^1(M)} / {\Omega_{\mathbb Z}^1(M)})$ for which we have the important result:
\bea
\label{trivstar}
\forall (\omega_0,\chi) \in \Omega_0^1(M) \times \Omega^1(M) \, , \, \, \, \, \, \, \, \, \,   \bar{\omega}_0 \star \bar{\chi} = \bar{\mathbf{0}} \, .
\eea

2) Let $(0 , d_{-1}(m_{\alpha \beta}/p_i), \tau_{\alpha \beta \gamma})$ and $(0 , d_{-1}(n_{\alpha \beta}/p_i), \kappa_{\alpha \beta \gamma})$ be representatives of two torsion origins $\bar{\mathbf{A}}_\tau$ and $\bar{\mathbf{A}}_\kappa$ of the configuration space. Their DB product admits as representative $(0 , 0 , 0, \tau_{\alpha \beta \gamma} \cdot d_{-1}(m_{\alpha \beta}/p_i), \tau_{\alpha \beta \gamma} \cdot \kappa_{\gamma \rho \epsilon})$ where the collection $(\tau_{\alpha \beta \gamma} \cdot \kappa_{\gamma \rho \epsilon})$ represent the cup product $\kappa \Cupp \tau$ and the collection $(\tau_{\alpha \beta \gamma} \cdot d_{-1}(m_{\alpha \beta}/p_i))$ the cup product $(\kappa \Cupp \tau) / p$ which on its turn defines the \emph{linking} of the torsion cycles $\kappa$ and $\tau$.

3)  Let $(0 , d_{-1}(m_{\alpha \beta}/p_i), \tau_{\alpha \beta \gamma})$ be a representatives of the torsion origin $\bar{\mathbf{A}}_\tau$ then definition (\ref{prodrep}) implies that:
\bea
\label{trivstar2}
\forall \omega \in \Omega^1(M) \, , \, \, \, \, \, \, \, \, \,   \bar{\mathbf{A}}_\tau \star \bar{\omega} = \bar{\mathbf{0}} \, .
\eea

\medskip

\noindent (2) {\it Generalized holonomy.} In order to integrate $\bar{\mathbf{A}} \star \bar{\mathbf{B}}$ over $M$ (seen as a $3$-cycle), we consider a polyhedral decomposition $(M_\alpha , S_{\alpha \beta} , L_{\alpha \beta \gamma} , x_{\alpha \beta \gamma \rho})$ of $M$. Mimicking the construction which led to definition (\ref{2.43}) we set:
\bea
\label{2.45}
\int_M \bar{\mathbf{A}} \star \bar{\mathbf{B}} \egzz  \sum_\alpha \int_{M_\alpha} A_\alpha \wedge d B_\alpha - \sum_{\alpha , \beta} \int_{S_{\alpha \beta}} \Lambda_{\alpha \beta} \wedge d B_\beta \, + \ \ \ \ \ \ \ \ \ \ \ \ \ \ \ \ \ \ \ \ \ \ \\ \nonumber
\sum_{\alpha , \beta , \gamma} \int_{L_{\alpha \beta \gamma}} m_{\alpha \beta \gamma} \wedge B_\gamma \, - \sum_{\alpha , \beta , \gamma , \rho} \int_{x_{\alpha \beta \gamma \rho}}  m_{\alpha \beta \gamma} \wedge \Theta_{\gamma \rho} \, .
\eea
Here also it can be checked that modulo integers this expression is independent of the representative of $\bar{\mathbf{A}}$ as well as of the polyhedral decomposition of $M$. Moreover for each $U_\alpha$ we recognize in the first term of expression (\ref{2.45}) the usual action of the $U(1)$ BF Quantum Field Theory in $U_\alpha \simeq \mathbb{R}^3$.

\vspace{2mm}

\noindent {\bf Examples:} 1) If we assumes that the DB classes $\bar{\mathbf{A}}$ and $\bar{\mathbf{B}}$ belong to the trivial fiber (the fiber over $\vec{0}$) then it can be shown that (\ref{2.45}) reduces to:
\bea
\label{2.45simple}
\int_M \bar{\mathbf{A}} \star \bar{\mathbf{B}} \egzz  \int_{M} \omega \wedge d \chi \, ,
\eea
where $\omega$ and $\chi$ are two 1-forms on $M$ such that $\bar{\mathbf{A}} = \bar{0} + \bar{\omega}$ and $\bar{\mathbf{B}} = \bar{0} + \bar{\chi}$. In particular, since $H_D^1(S^3,\mathbb{Z})$ has only a trivial fiber, expression (\ref{2.45simple}) becomes the generic one on $S^3$ (seen as a compactification of $\mathbb{R}^3$).

2) For a torsion origin $\bar{\mathbf{A}}_\tau$ with representative $(0 , d_{-1}(\zeta_{\alpha \beta}/p_i), \tau_{\alpha \beta \gamma})$ we have:
\bea
\label{squaretor}
\int_M \bar{\mathbf{A}}_\tau \star \bar{\mathbf{A}}_\tau \egzz - \sum_{\alpha , \beta , \gamma , \rho} \int_{x_{\alpha \beta \gamma \rho}}  \tau_{\alpha \beta \gamma} \cdot d_{-1}(\zeta_{\alpha \beta}/p_i) = - {{\tau \pitchfork \zeta} \over p_i} \egzz - Q(\tau , \tau)\, ,
\eea
with $p_i \cdot \zeta = \tau$, $\pitchfork$ denoting transverse intersection and $Q$ being the symmetric bilinear linking form on $T_1(M)$ \cite{Wa,Gr}. Strictly speaking, the integral in the right-hand side of the first equality is $\frac{1}{p_i} < \! \tau \Cupp \zeta \, , M >$ which through Poincaré duality is nothing but $\frac{1}{p_i}{\tau \pitchfork \zeta} $.

\medskip

\noindent (3) {\it Lagrangian, action and functional measure.} From relation (\ref{2.45simple}) and remark that follows it seems natural to define the $U(1)$ BF Lagrangian on a generic oriented closed smooth $3$-manifold $M$ as:
\begin{equation}
\label{bf1}
\mathfrak{bf}_1(\bar{\mathbf{A}},\bar{\mathbf{B}}) := \bar{\mathbf{A}} \star \bar{\mathbf{B}} \, ,
\end{equation}
with $(\bar{\mathbf{A}},\bar{\mathbf{B}}) \in H^1_D(M,\mathbb{Z}) \times H^1_D(M,\mathbb{Z})$, and in the presence of a coupling constant $k$ as:
\begin{equation}
\label{bfk}
\mathfrak{bf}_k(\bar{\mathbf{A}},\bar{\mathbf{B}}) := k \; \bar{\mathbf{A}} \star \bar{\mathbf{B}} \, .
\end{equation}
Since $H_D^3(M,\mathbb{Z}) \simeq \mathbb{R}/\mathbb{Z}$ expression (\ref{bfk}) is well-defined if and only if:
\bea
\label{quantizek}
k \in \mathbb{Z} \, .
\eea
In other words the {\bf coupling constant is quantized} as stated in Lemma 1. The corresponding action (with coupling constant $k$) is then:
\begin{equation}
\label{bfaction}
BF_k(\bar{\mathbf{A}},\bar{\mathbf{B}}) := \int_M \mathfrak{bf}_k(\bar{\mathbf{A}},\bar{\mathbf{B}}) = k \int_M {\bar{\mathbf{A}} \star \bar{\mathbf{B}}} \, .
\end{equation}
This is a $\mathbb{R}/\mathbb{Z}$-valued symmetric bilinear mapping on $H^1_D(M,\mathbb{Z}) \times H^1_D(M,\mathbb{Z})$. To simplify notations we will write this action as $BF_k$. Note that it would be more appropriate to call the theory "$U(1)$ AB" than "$U(1)$ BF".

The (formal) functional measure of $U(1)$ BF theory is then:
\begin{equation}
\label{8bf}
\mu_{BF_k} := \mu_{BF_k}(\bar{\mathbf{A}},\bar{\mathbf{B}}) := D\bar{\mathbf{A}} \; D\bar{\mathbf{B}} \cdot \exp\left\{
2i\pi k \int_M  {\bar{\mathbf{A}} \star \bar{\mathbf{B}} } \right\} \, .
\end{equation}
Assuming that we have picked up a full set of origins ${\bar{\mathbf{A}}}_{\vec{m}}$ on the fibers of $H_D^1(M,\mathbb{Z})$ (or of $H_D^1(M,\mathbb{Z})^\ast$) the functional measure can be decomposed according to:
\begin{equation}
\label{bfmeasure}
\mu_{BF_k} = \sum_{\vec{m},\vec{n} \in H^2(M)} D\overline{\alpha} \;  D\overline{\beta} \cdot \exp\left\{{2i\pi k \int_M \left( ({\bar{\mathbf{A}}}_{\vec{m}} + \overline{\alpha}) \star ({\bar{\mathbf{A}}}_{\vec{n}} + \overline{\beta}) \right) } \right\} \, ,
\end{equation}
where $D\overline{\alpha}$ (and $D\overline{\beta}$) is the (formal) functional Lebesgue measure on ${\Omega^1(M) / {\Omega_{\mathbb{Z}}^1(M)}}$ (or on $\left( \Omega_{\mathbb Z}^2(M) \right)^*$). Since $\bar{\mathbf{A}}$ and $\bar{\mathbf{B}}$ both belong to the same configuration space without any loss of generality we considered the same set of origins for each of them. Note that instead of summing over $H^2(M)$ we can sum over $H_1(M)$ as Poincaré duality implies that these two spaces are isomorphic.

Consider $\bar{\mathbf{a}},\bar{\mathbf{b}} \in H^1_D(M,\mathbb{Z})$ such that $ \bar{\mathbf{a}} \star \bar{\mathbf{b}}  = \bar{\mathbf{0}}$ and $k \bar{\mathbf{a}} = \bar{\mathbf{0}} = k \bar{\mathbf{b}}$. Then we have:
\bea
\label{periodBF}
BF_k(\bar{\mathbf{A}} + \bar{\mathbf{a}} , \bar{\mathbf{B}} + \bar{\mathbf{b}}) &=& k \int_M ( \bar{\mathbf{A}} \star \bar{\mathbf{B}} + \bar{\mathbf{A}} \star \bar{\mathbf{b}} + \bar{\mathbf{a}} \star \bar{\mathbf{B}} + \bar{\mathbf{a}} \star \bar{\mathbf{b}}) \nonumber \\
&=& k \int_M \bar{\mathbf{A}} \star \bar{\mathbf{B}} + \int_M \bar{\mathbf{A}} \star k \bar{\mathbf{b}} + \int_M k \bar{\mathbf{a}} \star \bar{\mathbf{B}} \; ,\\
&=& BF_k(\bar{\mathbf{A}} , \bar{\mathbf{B}}) \nonumber
\eea
and we say that the $U(1)$ BF theory is $k$-{\bf periodic}. As an example consider two closed 1-forms with integral periods, $\omega_1$ and $\omega_1$, and form the DB cocycles  $\mathbf{a} = (\omega_1|_{U_{\alpha}}/k , 0, 0)$ and $\mathbf{b} = (\omega_2|_{U_{\alpha}}/k , 0, 0)$. Then $ \bar{\mathbf{a}} \star \bar{\mathbf{b}}  = \bar{\mathbf{0}}$ and $k \bar{\mathbf{a}} = \bar{\mathbf{0}} = k \bar{\mathbf{b}}$.

Some remarks can be made concerning the use of distributional configuration spaces. If we consider $H^1_D(M,\mathbb{Z})^*$ as configuration space then we have to extend the DB product $\star$ to this space. This cannot be done straightforwardly because of the "product of distributions" plague. Nevertheless the DB product can be extended to a pairing on $H^1_D(M,\mathbb{Z}) \times H^1_D(M,\mathbb{Z})^*$ by setting: ${\bar{\mathbf{A}}} \star {\bar{\mathbf{\Phi}}} \leftrightarrow {\bar{\mathbf{\Phi}}}\left( {\bar{\mathbf{A}}} \right)$. If we consider as configuration space the simpler one $Z_1(M) \oplus H^1_D(M,\mathbb{Z})$ then the only problem in extending the DB product to this configuration space will be to define the DB product of $1$-cycles. It is not hard to see that for two homologically trivial $1$-cycles $z_1 = b \Sigma_1$ and $z_2 = b\Sigma_2$ with no points in common then $z_1 \star z_2$ is well-defined as the transverse intersection $\Sigma_1 \pitchfork z_2$ and hence is zero in $H_D^3(M,\mathbb{Z}) \simeq \mathbb{R}/\mathbb{Z}$. We then extend this procedure to all $1$-cycles with no common point, which defines the zero regularization procedure as explained in full details in \cite{GT1,T2}. This also coincides with the framing regularization usually used in CS Quantum Field Theory \cite{GMM,Gu}.
\vspace{2mm}

The Lagrangian for the $U(1)$ Chern-Simons theory on $S^3$ is $A \wedge dA$ and hence the same remarks as for the $U(1)$ Lagrangian hold. Accordingly on a generic smooth oriented closed $3$-manifold $M$ it seems natural to set:
\begin{equation}
\label{5}
\mathfrak{cs}_k({\bar{\mathbf{A}}}) := k {\bar{\mathbf{A}}} \star {\bar{\mathbf{A}}} = \mathfrak{bf}_k\left({\bar{\mathbf{A}}},{\bar{\mathbf{A}}}\right) \, ,
\end{equation}
which yields the $U(1)$ CS action with coupling constant $k$:
\begin{equation}
\label{7}
CS_k({\bar{\mathbf{A}}}) := \int_M {\mathfrak{cs}_k({\bar{\mathbf{A}}})}
= k \int_M {{\bar{\mathbf{A}}} \star {\bar{\mathbf{A}}}} \, .
\end{equation}
This seems to be a hint that a relation between the two models exists. However the $U(1)$ CS measure is
\begin{equation}
\label{8}
\mu_{CS_k} := \mu_{CS_k}({\bar{\mathbf{A}}}) := D{\bar{\mathbf{A}}} \cdot \exp\left\{
{2i\pi k\int_M  {{\bar{\mathbf{A}}} \star {\bar{\mathbf{A}}} } } \right\} \, .
\end{equation}
We still have to deal with the problem of product of distributions if we consider $H_D^1(M,\mathbb{Z})^\ast$ as configuration space instead of $H_D^1(M,\mathbb{Z})$. Assuming the same choices of origins as in the BF case, we can finally write:
\begin{equation}
\label{csmeasure}
\mu_{CS_k} = \sum_{\vec{n} \in H^2(M)} D\overline{\omega} \cdot \exp\left\{{2i\pi k \int_M  ({\bar{\mathbf{A}}}_{\vec{n}} + \overline{\omega}) \star ({\bar{\mathbf{A}}}_{\vec{n}} + \overline{\omega}) } \right\} \, .
\end{equation}
The difference between the two models now appears since in $\mu_{CS_k}$ there is only one family of integration parameters whereas in $\mu_{BF_k}$ there are two, one for each $H_D^1(M,\mathbb{Z})$ appearing in the configuration space. This will play a crucial role in the comparison of the two models.

As in the BF case we consider $\bar{\mathbf{a}} \in H^1_D(M,\mathbb{Z})$ such that $ \bar{\mathbf{a}} \star \bar{\mathbf{a}}  = \bar{\mathbf{0}}$ and $2k \bar{\mathbf{a}} = \bar{\mathbf{0}}$. We have:
\bea
\label{periodCS}
CS_k(\bar{\mathbf{A}} + \bar{\mathbf{a}}) &=& k \int_M ( \bar{\mathbf{A}} \star \bar{\mathbf{A}} + \bar{\mathbf{A}} \star \bar{\mathbf{a}} + \bar{\mathbf{a}} \star \bar{\mathbf{A}} + \bar{\mathbf{a}} \star \bar{\mathbf{a}} ) \nonumber \\
&=& k \int_M \bar{\mathbf{A}} \star \bar{\mathbf{A}} + \int_M \bar{\mathbf{A}} \star 2k \bar{\mathbf{a}} \; , \\
&=& CS_k(\bar{\mathbf{A}}) \nonumber
\eea
and we say that the CS theory is $2k$-{\bf periodic}. This periodicity property is particularly useful in order to show that homologically non-trivial free Wilson loops have a vanishing expectation value \cite{GT1,T2}. This complete the proof of part 1) of Lemma 2.

\subsection{Partition functions and 3-manifolds invariants}

\noindent As already mentioned at the level of actions the CS and BF theories are trivially related since:
\begin{equation}
\label{bfvscs}
CS_k({\bar{\mathbf{A}}}) = BF_k({\bar{\mathbf{A}}},{\bar{\mathbf{A}}}) \, .
\end{equation}
In other words $CS_k$ is the quadratic form associated with the symmetric bilinear form $BF_k$. Although on functional measures things become more tricky we can try to compare the two theories at the level of their partition functions.

\medskip

\noindent (1) {\it Definition of partition functions.} The partition function of the $U(1)$ CS theory is defined as:
\begin{equation}
\label{cspartition}
Z_{CS_{k}}(M) := \sum_{\vec{n} \in H_1(M)} { \int D\overline{\omega} \cdot \exp \left\{ {2i\pi k \int_M \left( ({\bar{\mathbf{A}}}_{\vec{n}} + \overline{\omega}) \star ({\bar{\mathbf{A}}}_{\vec{n}} + \overline{\omega}) \right) } \right\} \over \int D\overline{\omega} \cdot \exp\left\{{2i\pi k \int_M \overline{\omega} \star \overline{\omega}  } \right\} } \, .
\end{equation}
The computation can be straightforwardly adapted from the one of the BF parititon function which will be done later hence we simply give the final expression:
\begin{equation}
\label{cspart}
Z_{CS_{k}}(M)  = \sum_{\kappa_1=0}^{p_1-1} \sum_{\kappa_2=0}^{p_2-1} \cdots  \sum_{\kappa_d=0}^{p_d-1} e^{ 2 \pi i k \sum_{ij} \kappa_i \kappa_j Q_{ij} } = \sum_{\vec{\kappa} \in T_1(M)}  e^{ 2 \pi i k Q(\vec{\kappa},\vec{\kappa})}  \, .
\end{equation}
In this expression $T_1(M) = \mathbb{Z}_{p_1} \oplus \ldots \oplus \mathbb{Z}_{p_d}$ is the standard decomposition over abelian finite groups of the torsion part of the first homology group of $M$ (hence $p_i | p_{i+1}$) and $(Q_{ij})$ is the matrix of the non-singular symmetric bilinear linking form $Q: T_1(M) \times T_1(M) \longrightarrow {\mathbb{Q}/\mathbb{Z}}$. It is remarkable that only the torsion sector gives a non-trivial contribution to the partition function. This is precisely the difference between the partition function $Z_{CS_{k}}(M)$ and the Reshetikhin-Turaev invariant \cite{GT3,T2}. Note that the normalization of the partition function (\ref{cspartition}) differs from the one of \cite{DW}.

We then define the partition function of the $U(1)$ BF model as:
\begin{equation}
\label{bfpartition}
Z_{BF_{k}}(M) \equiv {\sum_{\vec{m},\vec{n} \in H_1(M)}} { \int \int D\overline{\alpha} \;  D\overline{\beta} \cdot \exp\left\{{2i\pi k \int_M \left( ({\bar{\mathbf{A}}}_{\vec{m}} + \overline{\alpha}) \star ({\bar{\mathbf{A}}}_{\vec{n}} + \overline{\beta}) \right) } \right\} \over \int \int D\overline{\alpha} \; D\overline{\beta} \cdot \exp\left\{{2i\pi k \int_M \overline{\alpha} \star \overline{\beta}  } \right\} } \, .
\end{equation}
There are arguments, made in the non-abelian case \cite{CCFM}, which indicate that the partition function of the BF model is related with the square norm of the CS one. Strictly speaking it is the non-abelian BF with cosmological model which is related, through the cosmological term, to the CS model. However in the abelian case the cosmological term is necessarily zero. Nevertheless the question about a possible relation between the two models remains interesting in the abelian case. Of course this also recalls the relation between the Turaev-Viro and the absolute square of the Reshetikhin-Turaev invariant, still in the non-abelian framework. In order to investigate this possible relation in the $U(1)$ case we must first consider $|Z_{CS_{k}}(M)|^2 = \overline{Z_{CS_{k}}(M)} Z_{CS_{k}}(M)$.

Concerning the infinite dimensional integration over ${\Omega ^1\left( M \right)} \mathord{\left/ {\vphantom {{\Omega ^1\left( M \right)} {\Omega _{\mathbb Z}^1 \left( M \right)}}} \right. \kern-\nulldelimiterspace} {\Omega _{\mathbb Z}^1 \left( M \right)}$ we can naively write:
\begin{equation}
\label{squaretranslation}
\left| \int D\overline{\alpha} \cdot e^{2i\pi k \int_M \overline{\alpha} \star \overline{\alpha} } \right|^2  = \int \int D\overline{\alpha}  D\overline{\beta} \cdot e^{2i\pi k \int_M (\overline{\alpha} \star \overline{\alpha} - \overline{\beta} \star \overline{\beta}) }  \, ,
\end{equation}
which after having set $\overline{\chi} = (\overline{\alpha} + \overline{\beta})$ and $\overline{\zeta} = (\overline{\alpha} - \overline{\beta})$ gives:
\begin{equation}
\label{squarebis}
\left| \int D\overline{\alpha} \cdot e^{2i\pi k \int_M \overline{\alpha} \star \overline{\alpha} } \right|^2  = \int \int D\overline{\chi}  D\overline{\zeta} \cdot e^{2i\pi k \int_M \overline{\chi} \star \overline{\zeta}  }  \, .
\end{equation}
However in the right-hand side of (\ref{squarebis}) integration is performed over the whole quotient ${\Omega^1(M) / {\Omega_{\mathbb{Z}}^1(M)}}$ whereas in the left-hand side integration is performed on the subspaces of ${\Omega^1(M) / {\Omega_{\mathbb{Z}}^1(M)}}$ generated by $\overline{\chi}$ and $\overline{\zeta}$. These subspaces do not coincide in general with ${\Omega^1(M) / {\Omega_{\mathbb{Z}}^1(M)}}$. Indeed the formal inversion of the relations defining $\overline{\chi}$ and $\overline{\zeta}$ in term of $\overline{\alpha}$ and $\overline{\beta}$ gives:
\bea
\label{inverse}
\left\{ \begin{gathered}
\overline{\alpha} = \frac{1}{2} (\overline{\chi} + \zeta) \\
\overline{\beta} = \frac{1}{2} (\overline{\chi} - \zeta)
\end{gathered} \right.
\, ,
\eea
expressions which are meaningless with respect to the quotient of $\Omega^1(M)$ by $\Omega^1_{\mathbb{Z}}(M)$ since $1/2$ times a closed form with integral periods is closed but in general not with integral periods. Nonetheless when the first de Rham cohomology group of $M$ is trivial then relations (\ref{inverse}) become meaningful. This happens for instance with $S^3$ or any lens space $L(p,q)$ with $p\geq2$. Quantum Field Theory is dealing with $S^3$ (actually rather with $\mathbb{R}^3$) this is why the subtlety remains unseen and leads to a trivial result: all partition functions are equal to 1.

\medskip

\noindent (2) {\it Taking zero-modes away.} In order to compute  partition function (\ref{bfpartition}) we concentrate on its numerator and more specifically on the argument of the exponential which yields the following four terms:
\bea
\label{decompargument}
{\bar{\mathbf{A}}}_{\vec{m}} \star {\bar{\mathbf{A}}}_{\vec{n}} \, +  \, {\bar{\mathbf{A}}}_{\vec{m}} \star \overline{\beta} \, + \, \overline{\alpha} \star {\bar{\mathbf{A}}}_{\vec{n}} \, + \, \overline{\alpha} \star \overline{\beta}
\eea
We write $\vec{m} = \vec{u} + \vec{\kappa}$ and $\vec{n} = \vec{v} + \vec{\tau}$, with $\vec{u}, \vec{v} \in F_1(M)$ and $\vec{\kappa}, \vec{\tau} \in T_1(M)$. Then the previous origins decompose according to:
\bea
\label{decomporigins}
{\bar{\mathbf{A}}}_{\vec{m}} &=& {\bar{\mathbf{A}}}_{\vec{u}} + {\bar{\mathbf{A}}}_{\vec{\kappa}} \\ \nonumber
{\bar{\mathbf{A}}}_{\vec{n}} &=& {\bar{\mathbf{A}}}_{\vec{v}} + {\bar{\mathbf{A}}}_{\vec{\tau}} \, .
\eea
Using property (\ref{trivstar2}) and commutativity of the DB product we can rewrite expression (\ref{decompargument}) as:
\bea
\label{decompargument2}
{\bar{\mathbf{A}}}_{\vec{u}} \star {\bar{\mathbf{A}}}_{\vec{v}} \, + \, {\bar{\mathbf{A}}}_{\vec{u}} \star {\bar{\mathbf{A}}}_{\vec{\tau}} \, + \, {\bar{\mathbf{A}}}_{\vec{v}} \star {\bar{\mathbf{A}}}_{\vec{\kappa}} \, + \, {\bar{\mathbf{A}}}_{\vec{\kappa}} \star {\bar{\mathbf{A}}}_{\vec{\tau}} \, + \, {\bar{\mathbf{A}}}_{\vec{u}} \star \bar{\beta} \, + \, \,{\bar{\mathbf{A}}}_{\vec{v}} \star \bar{\alpha} \, + \, \bar{\alpha} \star \bar{\beta}
\eea
There is an obvious exact sequence of abelian groups:
\bea
\label{exactOmega}
0 \longrightarrow {\left({\Omega_0^1(M) \over {\Omega_{\mathbb{Z}}^1(M)}}\right)} \xrightarrow{i_{can}} \left({\Omega^1(M) \over {\Omega_{\mathbb{Z}}^1(M)}}\right) \xrightarrow{j} \left({\Omega^1(M) \over {\Omega_0^1(M)}}\right) \longrightarrow 0 \, ,
\eea
from which we can give a meaning to the decomposition:
\bea
\label{decompOmega}
\left({\Omega^1(M) \over {\Omega_{\mathbb{Z}}^1(M)}}\right) \simeq \left({\Omega^1(M) \over {\Omega_0^1(M)}}\right) \times \left({\Omega_0^1(M) \over {\Omega_{\mathbb{Z}}^1(M)}}\right) = \left({\Omega^1(M) \over {\Omega_0^1(M)}}\right) \times \left({\mathbb{R} \over \mathbb{Z}}\right)^{b_1} \, .
\eea
First let $\rho^b$ be a set of closed $1$-forms on $M$ which dualize the previously chosen $1$-cycles $z_a$ which generate $F_1(M)$ and are free origins for the corresponding fiber of $H_D^1(M,\mathbb{Z})^\ast$:
\bea
\label{dualcycles}
\oint_{z_a} \rho^b = \delta_a^b \, .
\eea
For any set of angles $\theta_a \in \mathbb{R}/\mathbb{Z}$ let $\overline{\theta_a \rho^a}$ denotes the element of $(\Omega^1(M) / {\Omega_\mathbb{Z}^1(M)})$ defined by the closed $1$-form $\theta_a \rho^a$ (with Einstein's convention). It is clear from (\ref{exactOmega}) that:
\bea
j \left( \overline{\theta_a \rho^a} \right) = 0 \, .
\eea
Let $s:(\Omega^1(M) / {\Omega_0^1(M)}) \rightarrow \Omega^1(M)$ be a smooth section. This means that if $\, \, \, \widehat{} \, \, \,$ denotes the switch to the quotient $\Omega^1(M) / {\Omega_0^1(M)})$ then $\widehat{s(\widehat{\omega})} = \widehat{\omega}$ for any $\widehat{\omega} \in (\Omega^1(M) / {\Omega_0^1(M)})$. To any $\widehat{\omega} \in (\Omega^1(M) / {\Omega_0^1(M)})$ we associate the DB class $\overline{s\left( \widehat{\omega} \right)}$. Then for any set of angles $\theta_a \in \mathbb{R}/\mathbb{Z}$ we have:
\bea
j \left( \overline{s\left( \widehat{\omega} \right) + \theta_a \rho^a} \right) = j \left( \overline{s\left( \widehat{\omega} \right)} + \overline{\theta_a \rho^a} \right) =  j \left( \overline{s\left( \widehat{\omega} \right)} \right) = \widehat{\omega} \, .
\eea
By varying $\widehat{\omega}$ and the $\theta_a$'s we describe univocally the whole space $(\Omega^1(M) / {\Omega_\mathbb{Z}^1(M)})$ thus providing a meaning to decomposition (\ref{decompOmega}). This decomposition depends on the chosen section $s$. However the results we will obtain from this decomposition won't depend on $s$. Hence we adopt the simple notation $\widehat{\omega} + \overline{\theta_a \rho^a}$ without any reference to a section of $\Omega^1(M)$ over $(\Omega^1(M) / \Omega_0^1(M))$.

According to the now meaningful decomposition (\ref{decompOmega}) we write $\bar{\alpha} = \widehat{\alpha} + \overline{\theta_a \rho^a}$ and $\bar{\beta} = \widehat{\beta} + \overline{\varphi_a \rho^a}$ in such a way that, using property (\ref{trivstar}), the last three terms in expression (\ref{decompargument2}) are rewritten as:
\bea
\label{decompomega}
{\bar{\mathbf{A}}}_{\vec{u}} \star \overline{\theta_a \rho^a} \, + \, \,{\bar{\mathbf{A}}}_{\vec{v}} \star \overline{\varphi_a \rho^a} \, + \, {\bar{\mathbf{A}}}_{\vec{u}} \star \widehat{\beta} \, + \, \,{\bar{\mathbf{A}}}_{\vec{v}} \star \widehat{\alpha} \, + \, \widehat{\alpha} \star \widehat{\beta} \, .
\eea
This is the key of all the computation as we will now see. The part $D \bar{\alpha} D \bar{\beta}$ of the functional measure $\mu_{BF,k}$ decomposes on its turn according to (\ref{decompOmega}) as:
\bea
\left( D \widehat{\alpha} \cdot d^{b_1} \vec{\theta} \right) \left( D \widehat{\beta} \cdot d^{b_1} \vec{\varphi} \right) \, .
\eea
Integrations over the angles in the numerator of the BF partition function hence take the form:
\bea
\label{productfree}
\left( \int_{\left({\mathbb{R} \over \mathbb{Z}}\right)^{b_1}} d^{b_1} \vec{\theta} \, \, e^{2 i \pi k \int_M {\bar{\mathbf{A}}}_{\vec{u}} \star \overline{\theta_a \rho^a} } \right)
\left( \int_{\left({\mathbb{R} \over \mathbb{Z}}\right)^{b_1}} d^{b_1} \vec{\varphi} \, \, e^{2 i \pi k \int_M {\bar{\mathbf{A}}}_{\vec{v}} \star \overline{\varphi_a \rho^a} } \right) \, .
\eea
By construction, if we consider as configuration space $H_D^1(M,\mathbb{Z})^\ast$ then the free origins ${\bar{\mathbf{A}}}_{\vec{u}}$ and ${\bar{\mathbf{A}}}_{\vec{v}}$ are combinations of the $1$-cycles $z_a$. Hence we have:
\bea
\int_M {\bar{\mathbf{A}}}_{\vec{u}} \star \overline{\theta_a \rho^a} = \oint_{u^a z_a} \overline{\theta_b \rho^b} = u^a \theta_b \oint_{z_a} \rho^b = u^a \theta_a \, ,
\eea
where $u^a z_a$ is the $1$-cycle associated with ${\bar{\mathbf{A}}}_{\vec{u}}$. This reduces the factors in the product (\ref{productfree}) to delta symbols $\delta_{\vec{u} \, \vec{0}}$ and $\delta_{\vec{v} \, \vec{0}}$. This implies that the sum over $H_1(M)$ in the numerator of the BF partition function (\ref{bfpartition}) reduces to torsion sector and since we have chosen ${\bar{\mathbf{A}}}_{\vec{0}} = \bar{\mathbf{0}}$ the free contributions in (\ref{decompargument2}) all vanish. It then only remains:
\bea
\label{simplepartition}
\sum_{\vec{\kappa} , \vec{\tau}}  D \widehat{\alpha} \cdot D \widehat{\beta} \exp\left\{ 2 i \pi k \int_M \left(
{\bar{\mathbf{A}}}_{\vec{\kappa}} \star {\bar{\mathbf{A}}}_{\vec{\tau}} \, + \, \widehat{\alpha} \star \widehat{\beta} \right) \right\} \, .
\eea

\medskip

\noindent (3) {\it Partition functions as manifold invariants.} We can apply decomposition (\ref{decompOmega}) to the denominator of $Z_{BF_{k}}(M)$. The BF partition hence reduces to:
\bea
\label{BFpartitionend}
\sum_{\vec{\kappa} , \vec{\tau}}  D \widehat{\alpha} \cdot D \widehat{\beta} \exp \left\{ 2 i \pi k \int_M
{\bar{\mathbf{A}}}_{\vec{\kappa}} \star {\bar{\mathbf{A}}}_{\vec{\tau}} \right\} \, .
\eea
Finally from property (\ref{squaretor}) we conclude that:
\begin{equation}
\label{bfpartition2}
Z_{BF_{k}}(M)  =  \sum_{\vec{\kappa} \in T_1} \sum_{\vec{\tau} \in T_1}  e^{- 2 \pi i k Q(\vec{\kappa},\vec{\tau}) } \, .
\end{equation}
The explicit computation of this double sum is done in Appendix and gives:
\begin{equation}
\label{bfpartition3}
Z_{BF_{k}} = \prod\limits_{j=1}^{d}\gcd\left(k,p_{j}\right)p_{j} \, .
\end{equation}

In previous articles \cite{GT2,GT3,T2} it was shown that the partition function of the $U(1)$ CS theory on a smooth oriented closed $3$-manifold $M$ reads:
\begin{equation}
\label{cspart}
Z_{CS_{k}}(M)  =  \sum_{\vec{\kappa} \in T_1}  e^{- 2 \pi i k Q(\vec{\kappa},\vec{\kappa})}  \, .
\end{equation}
Hence as stated in part 2) of Lemma 2, only the torsion sector gives a non-trivial contribution to the partition function for the $U(1)$ CS and BF models. The comparison of expressions (\ref{cspart}) and (\ref{bfpartition2}) recalls the previous discussion concerning the infinite dimensional integrations over ${\Omega^1(M)} / {\Omega_{\mathbb Z}^1(M)}$ except that here we deal with torsion.

From expression (\ref{cspart}) we straightforwardly deduce that:
\begin{equation}
\label{cspartsq}
\left| Z_{CS_{k}}(M) \right|^2  =  \sum_{\vec{\kappa} \in T_1} \sum_{\vec{\tau} \in T_1}  e^{ 2 \pi i k (Q(\vec{\kappa},\vec{\kappa}) - Q(\vec{\tau},\vec{\tau}))}  \, .
\end{equation}
Details of the computation are given in Appendix and yield:
\begin{equation}
\label{cspartsq2}
\left| Z_{CS_{k}}(M) \right|^2  = 2^\gamma \left( \prod_{j=1}^d \; \gcd(k,p_j) \, p_j \right) \delta_{\beta 0} \, ,
\end{equation}
$\beta$ being the number of $p'_j$ such that $p'_j = 2(2l_j+1)$ and $\gamma$ the number of $p'_j$ such that $p'_j = 4l_j$, with $p'_j$ defined by $p_j = p'_j \gcd(k,p_j)$ ($j=1,\cdots,d$). Injecting relation (\ref{bfpartition3}) in expression (\ref{cspartsq2}) leads to:
\begin{equation}
\label{cs2vsbf}
\left| Z_{CS_{k}}(M) \right|^2  =  \delta_{\beta, 0} . 2^\gamma . Z_{BF_{k}}(M) = {2^{-\beta} \, \delta_{\beta, 0} \, Z_{BF_{2k}}(M)} \, ,
\end{equation}
which establishes the last part of Lemma 2. The second equality in (\ref{cs2vsbf}) can be derived from the reasoning done after equation (\ref{3.59}).

\vspace{2mm}
Before investigating in the next section the possible relation between the $U(1)$ CS and BF partition functions and a Turaev-Viro invariant of $M$, a final remark can be made. For a given coupling constant $k \in \mathbb{Z}$ the $U(1)$ CS partition function is connected to an abelian Reshetikhin-Turaev invariant as follows \cite{GT2,GT3}:
\begin{equation}
\label{RTvsCS}
\tau(M) = \frac{(2k)^{\slfrac{b_1}{2}}}{\sqrt{p_1 \ldots p_d}} \, Z_{CS_{k}}(M) \, .
\end{equation}
The construction of this RT invariant will be recalled in the last section. Furthermore if there is a $U(1)$ Turaev-Viro invariant which coincides with the absolute square of this  Reshetikhin-Turaev invariant (\ref{RTvsCS}) then relations (\ref{cs2vsbf}) implies that:
\begin{equation}
\label{TVvsCS2}
\Upsilon_{k}(M) = (2k)^{b_1} 2^\gamma \left( \prod_{j=1}^d \; \gcd(k,p_j) \right) \delta_{\beta 0} \, \, .
\end{equation}
Moreover, if there is a Turaev-Viro invariant related to the BF partition function then taking into account the same type of normalization as in equation (\ref{RTvsCS}) we should expect the equation:
\begin{equation}
\label{TVvsBF}
\Upsilon_{k}(M) = k^{b_1} \prod\limits_{j=1}^{d}\gcd\left(k,p_{j}\right) \, .
\end{equation}
The difference of periodicity of the two theories -- $2k$ for CS and $k$ for BF -- has been taken into account.

\section{A $U(1)$ Turaev-Viro invariant}

\noindent In this last section we shall try to relate the $U(1)$ CS and BF partition functions to some Reshetikhin-Turaev and Turaev-Viro invariants respectively. On the one hand it is well known that the non-abelian (typically $SU(2)$) CS partition function is a RT invariant \cite{Gu}. On the other hand it was proven in the context of modular categories that the absolute square of a RT invariant is a TV invariant \cite{T}. Still in the non-abelian case, it was formally shown that the BF partition function coincides with the absolute square of the CS parition function \cite{CCFM} thus showing that the non-abelian BF partition function is (up to some possible normalization) a TV invariant. This last result stems from connectedness of the set of classes of non-abelian (typically $SU(n)$) connections on a closed 3-manifold $M$. In the $U(1)$ case we saw that connectedness does not hold anymore. This is precisely what prevented us from doing the parametrization trick proposed after equation (\ref{squaretranslation}). Furthermore, it can be shown \cite{GT3} that the $U(1)$ CS partition function coincides, up to some  normalization, with an abelian RT invariant according to formula (\ref{RTvsCS}). Unfortunately we have seen that the relation between the $U(1)$ BF and CS partition functions is not as simple as in the non-abelian case since in general: $\left| Z_{CS_{k}}(M) \right|^2  \neq Z_{BF_{k}}(M)$. What we will show in this last section is that this inequality is explicitely due to the fact that the RT invariant related to the $U(1)$ CS theory is not coming from a modular category. We will show that if we relax the constraint that the RT invariant is related to the $U(1)$ CS theory the underlying category turns out to be modular and then the associated RT and TV invariants fulfill the usual theorem: $\left| Z_{CS_{k}}(M) \right|^2  = Z_{BF_{k}}(M)$.

\subsection{Semisimple, modular and spherical structures on $\mathbb{Z}_N$}

\noindent We consider the finite group $\mathbb{Z}_N$. It's Pontrjagin dual $\mathbb{Z}_N^*$, i.e. the set of its (one dimensional) irreducible representations, is a multiplicative finite group isomorphic to $\mathbb{Z}_N $.

\medskip

\noindent (1) {\it Monoidal structure.} The set $\mathbb{Z}_N^* $ can be turned into a {\bf monoidal} category $\mathbb{C}^{\mathbb{Z}_N}$ as follows. The objects of $\mathbb{C}^{\mathbb{Z}_N}$ are the elements $R_p$ of $\mathbb{Z}_N^*$, and its morphisms are the natural transformations in $\mathbb{Z}_N^*$, that is to say:
\bea
\label{decadix}
\Hom \, (R_p,R_q) = \delta_{p,q} \End \, \mathbb{C} \cong \delta_{p,q} \, \mathbb{C} = \delta_{p-q,0} \, \mathbb{C} \, ,
\eea
with an obvious convention and where the Kronecker delta is taken with respect to $\mathbb{Z}_N$. Note that composition in $\mathbb{C}^{\mathbb{Z}_N}$ is just multiplication in $\mathbb{C}$. In particular $\End \, (R_p) = \Hom \, (R_p,R_p) = \mathbb{C}$ and since this property defines simple objects we conclude that every object of $\mathbb{C}^{\mathbb{Z}_N}$ is simple.

The category $\mathbb{C}^{\mathbb{Z}_N}$ is endowed with the tensor product defined by:
\bea
\label{tensorprod}
R_p \otimes R_q = R_{p+q} \, ,
\eea
which trivially turns $\mathbb{C}^{\mathbb{Z}_N}$ into a monoidal category. The unit object is then $R_0$, that is to say the identity of $\mathbb{Z}_N^*$. Note that the tensor structure on $\mathbb{C}^{\mathbb{Z}_N}$ is canonically related to the addition in $\mathbb{Z}_N$ (and to the multiplication in $\mathbb{Z}_N^*$). In particular we have: $R_p \otimes R_q = R_q \otimes R_p$.

\medskip

\noindent (2) {\it Ribbon structure.} A braiding and a twist on $\mathbb{C}^{\mathbb{Z}_N}$ are isomorphisms $C$ and $\Theta$:
\bea
C_{p,q} : R_p \otimes R_q \rightarrow R_q \otimes R_p \; \; \; \; \; \; , \; \; \; \; \; \; \Theta_p : R_p \rightarrow R_p \, ,
\eea
which have to fulfill some constraints \cite{T}. Due to (\ref{tensorprod}) the braiding and the twist are actually defined by a set of complex numbers $\{c_{p,q},\theta_{p}\}$ such that these constraints simply reads:
\bea
\label{monoconstr}
c_{p,q+r} &=& c_{p,q} c_{p,r} \nonumber  \\
c_{p+r,q} &=& c_{p,q} c_{r,q} \; \; \; \; \; \; \; \; , \\
\theta_{p+q} &=& c_{q,p} c_{p,q} \theta_{p} \theta_{q} \nonumber
\eea
the Yang-Baxter equation being trivial in this abelian framework. The first two linearity conditions in (\ref{monoconstr}) straightforwardly yields:
\bea
\label{cpq}
c_{p,q} = (c_{1,1})^{pq} = c_{q,p}  \; \; \; \; \; \; , \; \; \; \; \; \; c_{p,0} = 1 = (c_{1,1})^{N} \, ,
\eea
and the last one:
\bea
\label{thetap}
\theta_{p} = (c_{1,1})^{p(p-1)} (\theta_{1})^p  \; \; \; \; \; \; , \; \; \; \; \; \; \theta_{0} = 1  \, .
\eea
The complex numbers $c_{p,q}$ are $N$-th roots of unity; typically $c_{1,1} = e^{2 i \pi /N}$. If $q$ and $r$ are two elements of $\mathbb{Z}_N$ such that $q + r = 0$ in $\mathbb{Z}_N$, then we have:
\bea
\label{invert}
c_{p,r} = (c_{p,q})^{-1} = (c_{1,1})^{-pq} = (c_{1,1})^{pr}  \; \; \; \; \; \; , \; \; \; \; \; \; \theta_{r} = (c_{1,1})^{q(q+1)} (\theta_{1})^{-q} \, .
\eea
Providing $\mathbb{C}^{\mathbb{Z}_N}$ with such a twist and braiding turns it into a {\bf ribbon} category which is denoted by $(\mathbb{C}^{\mathbb{Z}_N},c,\theta)$.

Duality in $\mathbb{C}^{\mathbb{Z}_N}$ is defined by setting:
\bea
\label{duality}
(R_p)^* = R_{N-p} \, .
\eea
Hence, like tensor product, duality is directly related to the (abelian) group structure of $\mathbb{Z}_N$. The trivial representation $R_0$ is obviously self-dual. There is another self-dual object if and only if $N=2k$ since then $(R_k)^* = R_{2k-k} = R_k$. From now on we denote $p^*$ the opposite of $p$ which also corresponds to the dual object $(R_p)^*$. This duality is compatible with the ribbon structure $(c,\theta)$ on $\mathbb{C}^{\mathbb{Z}_N}$ if
\bea
\label{compduality}
\theta_{p^*} = \theta_{p} \, .
\eea
Taking into account relation (\ref{invert}) this constraint gives:
\bea
\label{thetaduality}
(\theta_1)^{2p} = (c_{1,1})^{2p} \, ,
\eea
for any $p = 0, \cdots , N-1$ and hence:
\bea
\label{thetafixed}
\theta_1 = \epsilon \, c_{1,1} \, ,
\eea
where $\epsilon = \pm 1$. Finally we find that for $(\mathbb{C}^{\mathbb{Z}_N},c,\theta, \epsilon)$ to be a ribbon category we must have:
\bea
\label{ribbon}
c_{p,q} = (c_{1,1})^{pq} = (c_{p,q^*})^{-1} \; \; \; \; \; \; , \; \; \; \; \; \;
\theta_p = \epsilon^p (c_{1,1})^{p^2} = \epsilon^p c_{p,p} = \theta_{p^*} \, .
\eea

\medskip

\noindent (3) {\it S-matrix and modular structure.} Due to property (\ref{decadix}) we conclude that when constraints (\ref{ribbon}) are fulfilled $(\mathbb{C}^{\mathbb{Z}_N},c,\theta, \epsilon)$ is a {\bf semi-simple} category. In this category we introduce the trace operator:
\bea
\label{traceop}
tr(f_p) = c_{p,p^*} \, \theta_p \, f_p  \, ,
\eea
for any $f_p \in \End \, (R_p)$ as well as the dimension of an object $R_p$:
\bea
\label{dimp}
dim(p) = tr(Id_p) = c_{p,p^*} \, \theta_p  \, .
\eea
Thanks to relations (\ref{ribbon}) on gets:
\bea
\label{trace&dimo}
tr(f_p) = \epsilon^p \, f_p  \; \; \; \; \; \; , \; \; \; \; \; \;
dim(p) = \epsilon^p  \, .
\eea

The $S$-matrix of $(\mathbb{C}^{\mathbb{Z}_N},c,\theta, \epsilon)$ is defined as:
\bea
\label{Sexplicit}
S = \left( tr(c_{q,p}c_{p,q}) \right)_{p,q \in \mathbb{Z}_N} \, .
\eea
For the semi-simple category $(\mathbb{C}^{\mathbb{Z}_N},c,\theta, \epsilon)$ this gives:
\bea
\label{Smatrix}
S_{p,q} = \epsilon^{p+q} \, c_{q,p} \, c_{p,q} = \epsilon^{p+q} \, (c_{1,1})^{2pq} = S_{q,p} \, .
\eea
We can also write:
\bea
\label{S&twist}
S_{p,q} = \epsilon^{p+q} \, \theta_{p+q} \, (\theta_p)^{-1} \, (\theta_q)^{-1} \, ,
\eea
thus showing that the $S$-matrix is closely related with the bilinear form associated with the quadratic form defined by $\theta$. It is remarkable that in this abelian case the matrix elements of $S$ can be expressed with the braiding only. The dimension of an object $R_p$ can be expressed in terms of the $S$-matrix entries according to:
\bea
\label{Sdimp}
dim(p) = S_{p,0}  \, .
\eea

The semi-simple category $(\mathbb{C}^{\mathbb{Z}_N},c,\theta, \epsilon)$ is modular if and only if the $S$-matrix is invertible. We treat the case where $\epsilon = +1$ leaving to the reader the case $\epsilon = -1$ which is totally similar. It is not hard to check that $S$ has the form of a Wandermonde matrix:
\bea
\label{Wander}
S = \left( {\begin{array}{*{20}{c}}
  1&{{\alpha _0}}&{\alpha _0^2}& \cdots &{\alpha _0^{N - 1}} \\
  1&{{\alpha _1}}&{\alpha _1^2}& \cdots &{\alpha _1^{N - 1}} \\
  1&{{\alpha _2}}&{\alpha _2^2}& \cdots &{\alpha _2^{N - 1}} \\
   \vdots & \vdots & \vdots & \ddots & \vdots  \\
  1&{{\alpha _{N - 1}}}&{\alpha _{N - 1}^2}& \cdots &{\alpha _{N - 1}^{N - 1}}
\end{array}} \right)
\eea
with ${\alpha _p} = {\left( {{c_{1,1}}} \right)^{2p}}$. We then obtain:
\bea
\label{detWander}
\det S = \prod\limits_{1 \leqslant m < n \leqslant N} {\left( {{\alpha _n} - {\alpha _m}} \right)} \, .
\eea
Hence $S$ is not invertible if and only if at least two of the complex numbers ${\alpha _p}$ are equals. This means that:
\bea
\label{notinvert}
{\left( {{c_{1,1}}} \right)^{2m}} = {\left( {{c_{1,1}}} \right)^{2n}} \, ,
\eea
and then that:
\bea
\label{parity}
2(m-n) = 0 \quad \left[ N \right] \, .
\eea
Since the same reasoning will lead to the same result when $\epsilon = -1$, we find that $(\mathbb{C}^{\mathbb{Z}_N},c,\theta, \epsilon)$ is {\bf modular} if and only if $N = 2l+1$.

The tensor product and duality turns $\mathbb{C}^{\mathbb{Z}_{N}}$ into a so-called {\bf pivotal} category. It is easy to check that the left and right trace operators involved by this pivotal category are coinciding with the one of equation (\ref{trace&dimo}) thus turning $\mathbb{C}^{\mathbb{Z}_{N}}$ into a {\bf spherical} category. It is this spherical structure of the considered category which is involved in the construction of TV invariants of $M$ \cite{BW,BK}.

\subsection{The $\mathbb{Z}_N$ Reshetikhin-Turaev invariant versus the $U(1)$ Chern-Simons partition function}

\noindent It is the aim of this section to rephrase the Reshetikhin-Turaev construction of topological invariants obtained from surgery links in $S^3$ based on the category $\mathbb{C}^{\mathbb{Z}_{N}}$. Since any closed $3$-manifold can be obtained by performing a Dehn surgery on a link in $S^3$ this provides topological invariants for all closed $3$-manifolds. The details of the construction can be found in \cite{T}.

\medskip

\noindent (1) {\it Abelian RT invariant.} For the semi-simple category $(\mathbb{C}^{\mathbb{Z}_N},c,\theta, +1)$
we can associate to any framed link $\mathcal{L} = \bigcup_{i=1}^m \mathcal{L}_i$ with each component $\mathcal{L}_i$ holding a charge $p^i \in \mathbb{Z}_N$, the quantity:
\bea
\label{functorN}
F(\mathcal{L} , \vec{p}) = F(\mathcal{L} , p^1, \cdots , p^m) = \prod_{1 \leq p^i < p^j \leq m} \left( c_{p^j , p^i} \, c_{p^i , p^j} \right)^{L_{ij}} \prod_{p^i = 1}^{m} \left( c_{p^i , p^i} \right)^{L_{ii}} \, ,
\eea
where $L_{ij} = {\it lk}(\mathcal{L}_i,\mathcal{L}_j)$ define the linking matrix $\mathbf{L}$ of $\mathcal{L}$ with the convention that the self-linking number $L_{ii}$ is defined by the framing of $\mathcal{L}$. Taking into account equations (\ref{ribbon}) and (\ref{Smatrix}) (with $\epsilon = 1$) we get:
\bea
\label{functorN}
F(\mathcal{L} , \vec{p}) = \prod_{1 \leq p^i < p^j \leq m} \left( S_{p^i , p^j} \right)^{L_{ij}} \prod_{p^i = 1}^{m} \left( \theta_{p^i} \right)^{L_{ii}} = \left( c_{1,1} \right)^{\sum\limits_{i,j=1}^{m} p^i L_{ij} p^j} \, ,
\eea
and from the second relation of (\ref{cpq}) we conclude that:
\bea
\label{functorNbis}
F(\mathcal{L} , \vec{p})  = \exp \{ \frac{2 i \pi ({}^t\vec{p} \, \mathbf{L} \, \vec{p})}{N} \} \, ,
\eea
where ${}^t\vec{p}$ denotes the adjoint of $\vec{p}$. The corresponding {\bf RT invariant} is then defined as:
\bea
\label{RTgen}
\tau_N(M) = \Delta_N^{\sigma(\mathbf{L})} \mathcal{D}_N^{- \sigma(\mathbf{L}) - m - 1} \sum_{\vec{p} \in (\mathbb{Z}_N)^m} F(\mathcal{L} , \vec{p}) \, ,
\eea
where $\sigma(\mathbf{L})$ is the signature of $\mathbf{L}$, $\mathcal{D}_N^2 = N$ and $\Delta_N = \sum\limits_{p = 0}^{N-1} e^{-2 i \pi p^2/N}$. It turns out that $\Delta = 0$ if and only if $N = 2 (2k+1)$, $k \in \mathbb{Z}$. For $N \neq 2 (2k+1)$ we have:
\bea
\label{normDelta}
|\Delta_N|^2 = \sum\limits_{p,q = 0}^{N-1} e^{-2 i \pi (p^2 - q^2)/N} = N \, .
\eea
In this case, even if the category $\mathbb{C}^{\mathbb{Z}_N}$ might not be modular, formula (\ref{RTgen}) is a well-defined topological invariant of the $3$-manifold obtained by a Dehn surgery along $\mathcal{L}$ in $S^3$ \cite{T}.

As the sum in expression (\ref{RTgen}) only covers the $F$ term we can concentrate oneself on its properties. Remarkably when $N = 4k$ the sum has fundamental periodicity $2k$ and not $4k$ since:
\bea
\label{Fterm}
{}^t(\vec{p} +2k \vec{m}) \, \mathbf{L} \, (\vec{p} +2k \vec{m}) &=& {}^t\vec{p} \, \mathbf{L} \, \vec{p} + 2(2k) ({}^t\vec{p} \, \mathbf{L} \, \vec{m})) + (2k)^2 ({}^t\vec{m} \, \mathbf{L} \, \vec{p}) \nonumber \\
&=& {}^t\vec{p} \, \mathbf{L} \, \vec{p} \; \; \; \; [4k]  \, .
\eea
We can check that in any other case (i.e. if $N = 2k+1$ or $N = 2(2k+1)$) then the sum over $\mathbb{Z}_N$ in (\ref{RTgen}) is strictly $N$-periodic. Consequently when $N = 4k$ we have:
\bea
\label{reducedsum}
\sum_{\vec{p} \in (\mathbb{Z}_{4k})^m} F(\mathcal{L} , \vec{p}) = 2 \sum_{\vec{p} \in (\mathbb{Z}_{2k})^m} F(\mathcal{L} , \vec{p}) \,.
\eea
The same argument applies to $\Delta_{4k}$. We have $\mathcal{D}_{4k}^2 = 4k= 2(2k)$ and $\Delta_{4k} = 2 \sum\limits_{p = 0}^{2k-1} e^{-2 i \pi p^2/N} = 2 \, \Delta_{\frac{4k}{2}}$ and hence expression (\ref{RTgen}) takes the form:
\bea
\label{reducedRT}
\tau_{4k}(M) = \Delta_{\frac{4k}{2}}^{\sigma(\mathbf{L})} \, (\sqrt{2k})^{- \sigma(\mathbf{L}) - m} \sum_{\vec{p} \in (\mathbb{Z}_{2k})^m} F(\mathcal{L} , \vec{p}) \, .
\eea
Taking into account relation (\ref{normDelta}) we finally get:
\bea
\label{reducedRTfinal}
\tau_{4k}(M) = \left( \frac{\Delta_{\frac{4k}{2}}}{|\Delta_{\frac{4k}{2}}|} \right)^{\sigma(\mathbf{L})} \, |\Delta_{\frac{4k}{2}}|^{ - m} \sum_{\vec{p} \in (\mathbb{Z}_{2k})^m} e^{ 2 i \pi \frac{{}^t\vec{p} \, \mathbf{L} \, \vec{p}}{N} } \, .
\eea
This is precisely the invariant introduced by H. Murakami, T. Ohtsuki and M. Okada \cite{MOO} and which is an example of the most general family of abelian RT invariants introduced by F. Deloup \cite{Del}. We refer to expression (\ref{reducedRTfinal}) of $\tau_{4k}(M)$ as its {\it reduced expression}.

\medskip

\noindent (2) {\it Absolute square.} Once computed the absolute square of this invariant takes the following values \cite{MOO}:
\bea
\label{murakami2}
|\tau_{4k}(M)|^2 = \left\{ \begin{gathered}
 |H^1(M;\mathbb{Z}_{2k})| \; \; \; \; {\text{ if }} \; \alpha \Cupp \alpha \Cupp \alpha = 0 \; \; \forall \alpha \in H^1(M;\mathbb{Z}_{2k}) \hfill \\
 0 \; \; \; \; \; \; \; \; \; \; \; \; \; \; \; \; \; \; \; \; \; \;  \,  {\text{ otherwise }} \hfill \\
\end{gathered} \right. \, .
\eea
Note that the group appearing in this formula is $\mathbb{Z}_{2k}$ and not $\mathbb{Z}_{4k}$.

Instead of using $\alpha \Cupp \alpha \Cupp \alpha$ to differentiate the two cases appearing in (\ref{murakami2}) we can use Corollary 5.3 of \cite{MOO} which states that $|\tau_{4k}(M)|^2 = 0$ if and only if there exists $m>0$ and $\kappa \in H_1(M)$ of order $2^m$ such that $Q(\kappa,\kappa) = c/2^m$ and $k = 2^{m-1}b$, with $b$ and $c$ odd integers and $Q$ the linking form on $T_1(M)$. We introduce $p_i = p'_i \gcd(k,p_i)$ and $k = k'_i \gcd(k,p_i)$ ($i=1, \cdots, d$). On the one hand the $\delta_{\beta , 0}$ factor in expression (\ref{cs2vsbf}) tells us that $|Z_{CS_{k}}(M)|^2 = 0$ if and only if there exists (at least) one $p'_i$ with $p'_i = 2(2l_i+1)$. On the other hand $\gcd(k,p_i) = 2^n(2r+1)$ ($n\geq0$) and hence $p_i = 2^{n+1}(2r+1)(2l_i+1)$. Since $\gcd(k'_i,p'_i)=1$ we deduce that $k'_i = 2s_i+1$ and consequently that $k = 2^{n}(2r+1)(2s_i+1)$. By setting $m=n+1$, $\lambda = (2r+1)(2l_i+1)$ , $\kappa = \lambda \kappa_i$ ($\kappa_i$ a generator of $\mathbb{Z}_{p_i} \subset T_1$) and $b= (2r+1)(2s_i+1)$ we have $p_i = 2^{m} \lambda$, $2k= 2^m b$ and $Q(\kappa,\kappa) = \lambda^2 q_{ii}/p_i = \lambda \, q_{ii} / 2^{m}$ with $q_{ii}$ odd or the torsion order of $\kappa_i$ would be reduced. Finally if we set $c = \lambda \, q_{ii}$ (which is odd) we recover Corollary 5.3 of \cite{MOO}. The link between this reasoning and the cocycle $\alpha \Cupp \alpha \Cupp \alpha$ appearing in relation (\ref{murakami2}) will be shortly discussed at the end of the Appendix.

In order to study the case where $|\tau_{4k}(M)| \neq 0$ we first use the Universal Coefficient Theorem to write \cite{BoTu}:
\bea
\label{3.59}
H^1(M;\mathbb{Z}_{2k}) \cong \Hom\,(H_1(M),\mathbb{Z}_{2k}) \oplus \, \Ext(H_0(M),\mathbb{Z}_{2k}) \, ,
\eea
with $H_{1}(M) = F_{1}(M) \oplus T_{1}(M) = \mathbb{Z}^{b_{1}} \oplus \mathbb{Z}_{p_{1}} \oplus\cdots \oplus \mathbb{Z}_{p_{d}}$, and $H_{0}(M) = \mathbb{Z}$ (since $M$ is connected). Since $\Ext(\mathbb{Z},\mathbb{Z}_{2k}) = 0$ \cite{BoTu} the right-hand side of equation (\ref{3.59}) reduces to its first term. Furthermore as for abelian groups of finite type we have:
\bea
\label{3.60}
\Hom \left(\bigoplus_{i \in I} G_i,\mathbb{Z}_{2k}\right) = \bigoplus_{i \in I} \Hom\,(G_i,\mathbb{Z}_{2k}) \, ,
\eea
we simply have to determine the order of $\Hom\,(\mathbb{Z},\mathbb{Z}_{2k})$ and $\Hom\,(\mathbb{Z}_p,\mathbb{Z}_{2k})$ in order to determine $|\tau_{4k}(M)|^2$. The order of the first of these groups homomorphisms is $2k$ since $\Hom\,(\mathbb{Z},\mathbb{Z}_{2k}) \simeq \mathbb{Z}_{2k}$, and thus:
\bea
\label{3.61}
|\Hom\,(F_1(M),\mathbb{Z}_{2k})| = (2k)^{b_1} \, .
\eea
Since this will be of further interest in the sequel let us detail the computation of the order of $\Hom\,(\mathbb{Z}_p,\mathbb{Z}_{2k})$. For any $g \in \Hom\,(\mathbb{Z}_p,\mathbb{Z}_{2k})$ there exists $\nu_g \in \{0, \cdots , p-1\}$ such that:
\bea
 \left\{ \begin{gathered}
 g(0) = 0 \hfill \\
 g(n) = n.g(1) \hfill \\
 g(-n) \equiv g(p-n) = - g(n) = 2k \nu_g - g(n) \hfill
\end{gathered} \right. \, ,
\eea
for any $n \in \{0, \cdots , p-1\}$. Combining the second and the last of theses constraints we deduce that $(p-n).g(1) = 2k \nu_g - n.g(1)$ thus getting the constraint:
\bea
\label{3.63}
g(1) = \frac{2k \nu_g}{p} = \frac{2k' \nu_g}{p'} \, ,
\eea
with $p = p' \gcd(k,p)$ and $k = k' \gcd(k,p)$. As it is assumed that $|\tau_{4k}(M)| \neq 0$ the case where $p' = 2 (2l+1)$ has to be excluded and hence $p' = (2l+1)$ or $p' = 4l$. In the first case equation (\ref{3.63}) has $\gcd(k,p)$ solutions: $\nu_g \in \{0 , p',  \cdots , p-p'\}$ and $g(1) \in \{0, 2k', \cdots , 2(k- k')\}$. In the second case equation (\ref{3.63}) admits $2.\gcd(k,p)$ solutions: $\nu_g \in \{0 , p'/2, p',  \cdots , p-p'/2\}$ and $g(1) \in \{0, 2k', \cdots , (\gcd(k,p) - 1)2k'\}$. Combining all these results together we conclude that:
\bea
\label{orderfinal}
|H^1(M,\mathbb{Z}/2k\mathbb{Z})| = (2k)^{b_1} \, 2^\gamma \, \prod_{i=1}^d \gcd(k,p_i) \, ,
\eea
where $\gamma$ denotes the number of $p'_i$ such that $p' = 4l$. If $\beta$ is the number of $p'_i$ such that $p'_i = 2 (2l_i+1)$ then result (\ref{murakami2}) can be written under the alternate form:
\bea
\label{murakami3}
|\tau_{4k}(M)|^2 = (2k)^{b_1} \, \delta_{\beta,0} \, 2^\gamma \, \prod_{i=1}^d \gcd(k,p_i) \, .
\eea
Taking the normalization (\ref{RTvsCS}) into account we recover expression (\ref{cspartsq2}) of $|Z_{CS_{k}}(M)|^2$.

\vspace{2mm}

\noindent {\bf Examples:} 1) For $M = S^3$ we have $\tau_{4k}(M) = 1 = Z_{CS_{k}}(M)$ and hence $|\tau_{4k}(M)|^2 = 1 = |Z_{CS_{k}}(M)|^2$.

2) For $M = S^1 \times S^2$ we have $|\tau_{4k}(M)|^2 = 2k = (2k) \, |Z_{CS_{k}}(M)|^2$ since in that case $Z_{CS_{k}}(M) = 1$.

3) For the lens space $M = L(2,1) = \mathbb{R}P^3$ we have $|\tau_{4k}(M)|^2 = 2 = |Z_{CS_{k}}(M)|^2$ if $k$ is even and $|\tau_{4k}(M)|^2 = 0 = |Z_{CS_{k}}(M)|^2$ if $k$ is odd.

4) For a lens space $M = L(p,q)$ we find that neither $|\tau_{4k}(M)|^2$ nor $|Z_{CS_{k}}(M)|^2$ depends on $q$.

To complete this subsection let us consider the last two cases: $N=2(2k+1)$ and $N=2k+1$. It is only in the latter case that the category $\mathbb{C}^{\mathbb{Z}_N}$ is modular. We leave to the reader to prove that the value of the RT invariant associated with $\mathbb{Z}_{2k+1}$ is:
\bea
\label{tau2k+1}
|\tau_{2k+1}(M)|^2 = (2k+1)^{b_1} \, \prod_{i=1}^d \gcd(2k+1,p_i) = |H^1(M,\mathbb{Z}_{2k+1})| \, ,
\eea
(see Lemma 3.2 in \cite{MOO}). Finally, as already mentioned, there is no RT invariant when $N=2(2k+1)$. Furthermore there is no $U(1)$ CS theory corresponding to $N=2(2k+1)$ and $N=2k+1$. This completes the proof of the results concerning RT invariants in Lemma 3.

\vspace{2mm}

\noindent {\bf Example:} For a lens space $M = L(p,q)$ we have $|\tau_{2k+1}(M)|^2 = 1$ when $p$ is even (and not a multiple of $2k +1$).

\subsection{The $\mathbb{Z}_N$ Turaev-Viro invariant versus the $U(1)$ BF partition function}

\noindent In order to build an abelian TV invariant we follow the method initiated by S. Gelfand and D. Kazhdan \cite{GK} on the one hand and  J. Barrett and B. Westbury \cite{BW} on the other hand, and fully developed by B. Balsam and A. Kirillov \cite{BK}, rather than the original one of V. Turaev and O. Viro \cite{TV}. The advantage of this approach is that we can use any kind of polyhedral decomposition of an oriented closed $3$-manifold $M$ and not just simplicial ones. In order to have a chance to find a relation between TV and RT invariants we consider $\mathbb{C}^{\mathbb{Z}_N}$ which is a spherical category as explained in subsection 3.1.

\medskip

\noindent (1) {\it Oriented polyhedral decomposition.} Let $\Pi_M$ be a polyhedral decomposition of $M$ made of polyhedra (not necessarily tetrahedra), faces, edges and vertices. These objects are generically referred as $n$-cells of $\Pi_M$, with $n=0,1,2,3$ corresponding to vertices, edges, faces and polyhedra respectively. Strictly speaking a $n$-cell is homeomorphic to $\mathbb{R}^n$ and should not be confused with it closure. In the construction that follows only closure of cells are involved so when talking about a $n$-cell we always mean it closure.

Since $M$ is oriented all cells of $\Pi_M$ are orientable. If $C$ is a cell of $\Pi_M$ we denote by $C^o$ this cell once it is endowed with an orientation, and $-C^o$ the same cell but with the opposite orientation. However, thanks to Poincaré duality every $0$-cell (i.e. point) of $M$ has a canonical positive orientation. When writing a $0$-cell of $M$ as $x$ we always means this cell endowed with its canonical positive orientation whereas $x^o$ refers to either $x$ or $-x$. Furthermore, for $n=1,2,3$ any oriented $n$-cell $C^o$ induces an orientation for its bounding $(n-1)$-cells called the {\bf relative orientation}. More precisely, an oriented polyhedron $P^o$ provides its bounding faces with a canonical orientation via the "from inside to outside" rule; an oriented face $F^o$ provides a canonical orientation to its bounding edges by taking them in the counterclockwise order (a.k.a. Ampère's right hand screw rule); an oriented edge $\sigma^o$ provides a canonical orientation to its two bounding vertices (i.e. end points) via the "final minus initial" rule. Moreover, since $M$ is $3$-dimensional and closed any face $F$ of $\Pi_M$ is shared by exactly two oriented polyhedra $P^o_1$ and $P^o_2$ with respect to which $F$ is endowed with two orientations, $F^o_1$ and $F^o_2$, such that $F^o_2 = - F^o_1$. We respectively denote by $\mathcal{P}^o_\Pi$, $\mathcal{F}^o_\Pi$, $\mathcal{E}^o_\Pi$ and $\mathcal{V}^o_\Pi$ the set of oriented polyhedra, faces, edges and vertices of $\Pi_M$. These sets are provided with the standard structure of abelian free groups so that together with their boundary operator they give rise to a chain complex $K_\Pi$:
\bea
\label{complecPi}
\mathcal{P}^o_\Pi \mathop \to \limits^{\partial_3} \mathcal{F}^o_\Pi \mathop \to \limits^{\partial_2}  \mathcal{E}^o_\Pi \mathop \to \limits^{\partial_1}  \mathcal{V}^o_\Pi \, .
\eea
and to homology groups $H_i(\Pi_M)$. The polyhedral decomposition $\Pi_M$ is chosen in such a way that:
\bea
\label{gooddecomp}
H_i(\Pi_M) \cong H_i(M) \, ,
\eea
for $i=0,1,2,3$. We can use a good cover of $M$ to generate such a "good" polyhedral decomposition. This ensures its existence since $M$ always admit a good cover \cite{BoTu}.

\medskip

\noindent (2) {\it Labelings, state space and abelian TV invariant.} A {\bf $\mathbf{\mathbb{Z}_N}$-labeling} of $\Pi_M$ is a linear map $l : \mathcal{E}^{o}_\Pi \rightarrow  \mathbf{\mathbb{Z}_N}$ such that:
\bea
\label{duallabeling}
\hspace{-1cm} \forall \sigma^o \in \mathcal{E}^{o}_\Pi , \hspace{1cm} l(- \sigma^o) = l(\sigma^o)^* = N - l(\sigma^o) \, .
\eea
The $\mathbb{Z}_N$-valued number $l(\sigma^o)$ is a $\mathbb{Z}_N$-charge of $\sigma^o$. As there is no possible confusion we only refer to labelings and charges, without any more mention of the finite group $\mathbb{Z}_N$. The set of labeling of $\Pi_M$ is denoted by $\mathcal{L}_\Pi$.

Given a labeling $l$ of $\Pi_M$ we associate to every oriented face $F^o$ with bounding edges $\sigma^o_i$ the {\bf state space}:
\bea
\label{statespace}
H(F^o,l_\Pi) = \Hom\,(R_0,R_{l_1} \otimes \cdots \otimes R_{l_{n_{F^o}}})= \Hom\,(R_0,R_{\Sigma_{F^o}}) = \delta_{ \Sigma^l_{F^o} ,0 } \, \mathbb{C} \, ,
\eea
where $\Sigma^l_{F^o} = \sum\limits_{i=1}^{n_{F^o}} l_i = \sum\limits_{i=1}^{n_{F^o}} l(\sigma^o_i)$, the edges $\sigma^o_i$ being canonically oriented with respect to $F^o$. In this definition the Kronecker delta is taken in $\mathbb{Z}_N$ and the third equality derives from property (\ref{decadix}). Having $F^o$ running over the set of oriented faces of $\Pi_M$, the sums $\Sigma^l_{F^o}$ generate a linear map:
\bea
\Sigma^l : \mathcal{F}^{o}_\Pi \rightarrow \mathbb{Z}_N \, .
\eea

The {\bf total state space} of $\Pi_M$ with labeling $l$ is defined as:
\bea
\label{totstsp}
H(\Pi_M,l) = \bigotimes_{F}  \left( H(F^o_1,l) \otimes H(F^o_2,l) \right) = \prod_F \left(\delta_{\Sigma^l_{F^o_1} , 0} \, \delta_{ \Sigma^l_{F^o_2} , 0} \right)  \mathbb{C} \, ,
\eea
where $F$ runs over all unoriented faces of $\Pi_M$, $F^o_1$ and $F^o_2$ denoting $F$ endowed with its two possible orientations. However, as $\Hom\,(R_0,R_p)^* \cong \Hom\,(R_0,R^*_p)$ and $\delta_{N-p,0} = \delta_{p,0}$, we have:
\bea
\label{dualstatespace}
H(F^o,l)^* = H(-{F^o},l) = \delta_{N - \Sigma^l_{F^o} , 0} \, \mathbb{C} = \delta_{ \Sigma^l_{F^o} , 0} \, \mathbb{C} = H({F^o},l) \, ,
\eea
and hence the total state space of $\Pi_M$ for the spherical category $\mathbb{C}^{\mathbb{Z}_N}$ takes the simpler form:
\bea
\label{simpltotstsp}
H(\Pi_M,l) =  \left( \prod_F \delta_{ \Sigma^l_{F} , 0} \right)  \mathbb{C} \, ,
\eea
where $\Sigma^l_{F}$ is computed by using any of the two possible orientations of $F$. The trace operator in the spherical category $\mathbb{C}^{\mathbb{Z}_N}$ being the trace operator in $ \mathbb{C}$, the expression defining the $\mathbb{Z}_N$ {\bf TV invariant} of $M$ reduces to:
\bea
\label{TVab}
\Upsilon_{N}(M) = N^{-(v-1)} \sum_l \left( \prod_F \delta_{ \Sigma^l_{F} , 0} \right) \, ,
\eea
where $v$ is the number of (unoriented) vertices of $\Pi_M$. The normalization factor is usually taken to be $N^{-v}$ and not $N^{-(v-1)}$. However the relation with $BF$ partition function is simpler with the latter convention, and somehow more natural. In order to explicit this relation we need some complementary information.

\medskip

\noindent (3) {\it Cohomological computation.} A {\bf $\mathbf{\mathbb{Z}_N}$-gauging} of $\Pi_M$ is a linear map $\lambda : \mathcal{V}^{o}_\Pi \rightarrow  \mathbf{\mathbb{Z}_N}$ such that:
\bea
\label{dualgauging}
\hspace{-1cm} \forall x^o \in \mathcal{V}^{o}_\Pi , \hspace{1cm} \lambda(- x^o) = \lambda(x^o)^* = N - \lambda(x^o) \, .
\eea
The set of gauging of $\Pi_M$ is denoted by $\mathcal{G}_\Pi$. We can associate to any gauging $\lambda$ of $\Pi_M$ a labeling $d_0 \lambda$ of $\Pi_M$ by setting for any oriented edge $\sigma^o$:
\bea
\label{diffgauging}
\hspace{-1cm} \forall \sigma^o \in \mathcal{E}^o_\Pi , \hspace{1cm} \left( d_0 \lambda \right) (\sigma^o) = \lambda(\partial \sigma^o)  \, .
\eea
Denoting $x_1$ (resp. $x_2$) the initial (resp. final) point of $\sigma^o$, we can write:
\bea
\label{diffgauging2}
\left( d_0 \lambda \right) (\sigma^o) = \lambda(x_2) + \lambda(-x_1) = \lambda(x_2) - \lambda(x_1)  \, .
\eea
In particular we have $\left( d_0 \lambda \right) (- \sigma^o) = \lambda(x_1) - \lambda(x_2) = - \left( d_0 \lambda \right) (\sigma^o)$ as required. The map $d_0 : \mathcal{G}_\Pi \rightarrow \mathcal{L}_\Pi$ thus defined is called the {\bf differential} of $\mathcal{G}_\Pi$. A gauging $\lambda$ of $\Pi_M$ is said to be {\bf constant} if $d_0 \lambda = 0$. A constant gauging thus satisfies $\lambda(x_2) = \lambda(x_1)$ for the two end points of any oriented edge of $\Pi_M$. It is then obvious that:
\bea
\label{gaugecohom}
d_0 \lambda = 0 \Longleftrightarrow \left( \exists \lambda_0 \in \mathbb{Z}_N \, \, | \, \, \, \forall x \in M, \hspace{2mm} \lambda(x) = \lambda_0 \right) \, .
\eea

By construction the linear map $\Sigma^l : \mathcal{F}^{o}_\Pi \rightarrow \mathbb{Z}_N$ associates to each face a $\mathbb{Z}_N$-charges and thus define a {\bf $\mathbf{2}$-labeling} of $\Pi_M$ that is to say a linear map $f : \mathcal{F}^{o}_\Pi \rightarrow \mathbb{Z}_N$ such that $f(-F^o) = N - f(-F^o)$ for all oriented face $F^o$ of $\Pi_M$. Clearly $\Sigma^l$ is not the most general $2$-labeling but it derives from a labeling $l : \mathcal{E}^{o}_\Pi \rightarrow  \mathbf{\mathbb{Z}_N}$. Moreover for any gauging $\lambda$ of $\Pi_M$ definition (\ref{diffgauging}) yields:
\bea
\label{sumgauging}
\Sigma^{d_0 \lambda} = 0 \, .
\eea
Thus having $l$ running over $\mathcal{L}_\Pi$ the maps $\Sigma^l$ generate as a linear operator $d_1 : \mathcal{L}_\Pi \rightarrow \mathcal{C}_\Pi$, where $\mathcal{C}^o_\Pi$ denotes the set of $2$-labelings of $\Pi_M$, for which equation (\ref{sumgauging}) takes the form:
\bea
d_1 \circ d_0 = 0 \, ,
\eea
thus showing that $d_1$ is the differential of $\mathcal{L}_\Pi$. It is then straightforward to check that there is a notion of $3$-labelings of $\Pi_M$ and a linear operator $d_2$ on $\mathcal{C}_\Pi$ such that $d_2 \circ d_1 = 0$. This turns the collection of gaugings and $n$-labelings into a chain complex $K^*_\Pi$:
\bea
\label{complecPi}
\mathcal{G}_\Pi \mathop \to \limits^{d_0} \mathcal{L}_\Pi \mathop \to \limits^{d_1}  \mathcal{C}_\Pi \mathop \to \limits^{d_2}  \mathcal{M}_\Pi \, ,
\eea
where by definition $\mathcal{G}_\Pi = \Hom \,( \mathcal{V}^o_\Pi,\mathbb{Z}_N)$, $\mathcal{L}_\Pi = \Hom \,( \mathcal{E}^o_\Pi,\mathbb{Z}_N)$, $\mathcal{C}_\Pi = \Hom \,( \mathcal{F}^o_\Pi,\mathbb{Z}_N)$ and $\mathcal{M}_\Pi = \Hom \,( \mathcal{P}^o_\Pi,\mathbb{Z}_N)$. From the standard theory of homology and cohomology the homology of $K^*_\Pi$ defines the cohomology groups $H^i(\Pi_M,\mathbb{Z}_N)$ \cite{Do}. From property (\ref{gaugecohom}) we straightforwardly deduce that:
\bea
\label{constgaugecoho}
H^0(\Pi_M,\mathbb{Z}_N) = \mathbb{Z}_N \, = H^0(M,\mathbb{Z}_N) \, ,
\eea
in agreement with the Universal Coefficient Theorem, the latter furthermore involving that $H^1(\Pi_M,\mathbb{Z}_N) \simeq \Hom\,(H_1(\Pi_M),\mathbb{Z}_N) \oplus \Ext\,((H_0(\Pi_M),\mathbb{Z}_N))$. Since $\Pi_M$ is assumed to be good, relation (\ref{gooddecomp}) yields:
\bea
H^1(\Pi_M,\mathbb{Z}_N) \simeq \Hom\,(H_1(M),\mathbb{Z}_N) \oplus \Ext\,((H_0(\Pi_M),\mathbb{Z}_N)) = H^1(M,\mathbb{Z}_N) \, .
\eea

Returning to expression (\ref{TVab}) of $\Upsilon_{N}(M)$ we see that the non-zero contributions to the product over faces are those for which $\Sigma^l_{F^o} = 0$ for all oriented face $F^o$ of $\Pi_M$, in which case the contribution to $\Upsilon_{N}(M)$ is $1$. Thus we conclude that:
\bea
\left( \prod_F \delta_{ \Sigma^l_{F} , 0} = 1 \right) \Longleftrightarrow d_1 l = 0  \, .
\eea
Hence a labeling gives a non-trivial contribution to $\Upsilon_{N}(M)$ if and only if it is closed, the sum over labelings thus reducing to a sum over closed labelings. Unfortunately the set of closed labeling of $\Pi_M$ depends on $\Pi_M$ (see examples below) and hence doesn't define an invariant of $M$. Since a contributing labeling is a generator of the first homology group of the complex $K^*_\Pi$ and as $H^1(\Pi_M,\mathbb{Z}_N) = H^1(M,\mathbb{Z}_N)$ only depends on $M$ a natural solution would be to consider cohomology classes of closed labels. In other words we have to quotient the sum by the order of $\mathcal{G}^o_\Pi$. Since a gauging is endowing all points of $\Pi_M$ with a charges which are independent to each other we have:
\bea
\label{gaugingorder}
| \mathcal{G}^o_\Pi | = N^v \, .
\eea
Note that $d_0 \lambda$ is a contributing labeling for any gauging $\lambda$.
However, as constant gaugings do not change labelings ($l + d_0 \lambda_0 = l$), it seems more logical to quotient the sum over closed labelings not by the whole set $\mathcal{G}^o_\Pi$ but rather by $\mathcal{G}^o_\Pi / \mathcal{G}^o_{\Pi,0}$ where $\mathcal{G}^o_{\Pi,0}$ denotes the group of constant gaugings of $\Pi_M$. From equation (\ref{constgaugecoho}) we deduce that $\mathcal{G}^o_{\Pi,0} = \mathbb{Z}_N$ and hence that:
\bea
\label{gaugingorder}
| \mathcal{G}^o_\Pi / \mathcal{G}^o_{\Pi,0} | = N^{v-1} \, .
\eea
We obtain the final result:
\bea
\label{TVfinalresult}
\Upsilon_{N}(M) = |H^1(M,\mathbb{Z}_N)| \, ,
\eea
which completes the proof of Lemma 3. Note that the normalization $N^{-(v-1)}$ appearing in definition (\ref{TVab}) has now a cohomological interpretation.

\vspace{2mm}
\noindent {\bf Remark:} A polyhedral decomposition of $M$ induces a decomposition of the handle body $H_g$ of a Heegaard splitting of $M$. On its turn this decomposition induce two decomposition of the Riemann surface $\partial H_g$ related by the homeomorphism $h$ of $\partial H_g$ under the action of which the two copies of $H_g$ give rise to $M$. In other words we obtain two decomposition of $\partial H_g$ which are $h$-compatible on $\partial H_g$. The converse is obviously true. Hence by finding two decompositions of $H_g$ which are $h$-compatible on $\partial H_g$ we can compute the corresponding TV invariant of $M = H_g \cup_h H_g$. Note that due to result (\ref{TVfinalresult}) we only have to compute the first cohomology group of $M$. However it can be interesting to see how this definitions is working by itself in some simple cases.

\begin{figure}
  \centering
  \includegraphics[scale=0.35]{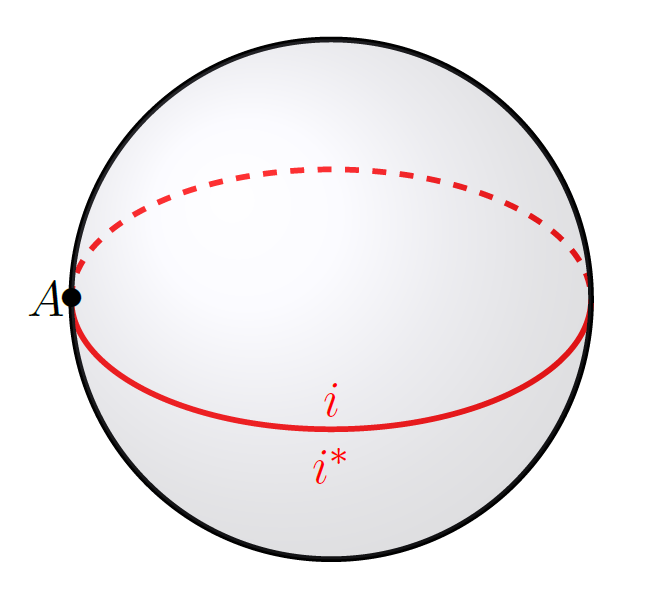}
  \caption{Polyhedral decomposition of $S^2$ compatible with the "basic" Heegaard splitting of $S^3$.}\label{fig4}
\end{figure}

\vspace{2mm}
\noindent {\bf Examples:} 1) The sphere $S^3$ can be obtained by gluing via the identity two $3$-balls (with opposite orientation) along their bounding $2$-sphere $S^2$. This Heegard splitting of $S^3$ is called "basic". Therefore in order to determine $\Upsilon_{N}(S^3)$ we just need to decompose $S^2$, the $3$-ball being identifiable with (the interior of) a polyhedron. We cut $S^2$ along an equator and add a vertex $A$ on this equator to get an edge the equator) and a vertex ($A$), and two faces (two hemispheres) as shown in Figure \ref{fig4}. By construction we set the charge $i$ on one hemisphere and $i^*$ on the other. Symmetrically the same decomposition holds on a second $S^2$ but with the charges $j$ and $j^*$. These two $S^2$ have to be glued together along their equator and vertices. Taking orientations into account this means that $i \leftrightarrow j^*$ and $i^* \leftrightarrow j$ for the gluing to yield a total state space of $M$.

The sum over labelings in the definition of $\Upsilon_{N}(S^3)$ gives:
\bea
\sum\limits_{i=0}^{N-1}\delta_{i,0}=1 \, ,
\eea
and the final expression of this TV invariant is:
\bea
\Upsilon_N\left(S_{3}\right) = \frac{1}{N^{1-1}} 1 = 1 \, .
\eea

\begin{figure}
  \centering
  \includegraphics[scale=0.5]{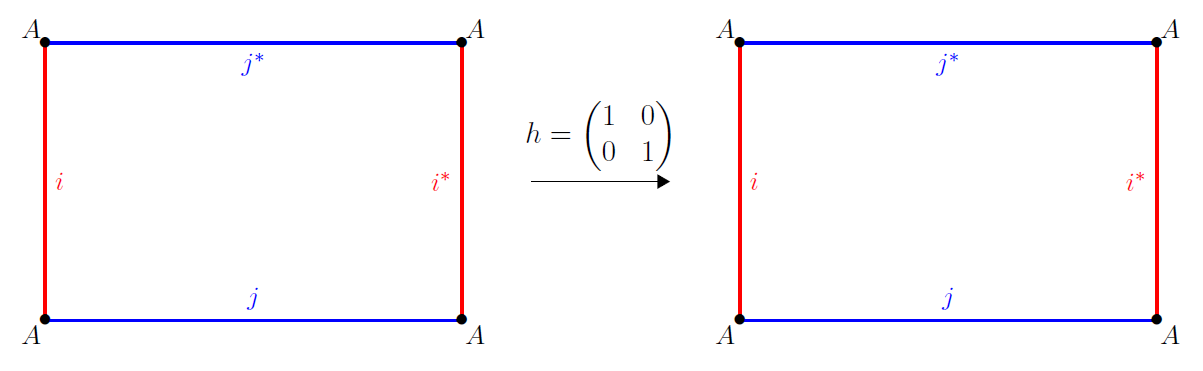}
  \caption{Polyhedral decomposition of $T^2$ compatible with the "basic" Heegaard splitting of $S^1 \times S^2$.}\label{fig5}
\end{figure}

2) The product manifold $S_{1} \times S_{2}$ can be obtained as any lens space, by gluing two solid tori along their boundary. A solid torus can be obtained by gluing together two non-intersecting disks, $D_1$ and $D_2$, on the boundary $S^2$ of a $3$-ball $B^3$. The boundary of a solid torus is $T^2 = S^1 \times S^1$ and is represented by a rectangle with opposite edges identified. One of the non-identified edges has to be the boundary of one of the disks, say $D_1$ of $B^3$. This edges and the one opposite to it on the rectangle are drawn in blue in Figure \ref{fig5}. We end with two edges (blue and red) and one vertex ($A$) on $T^2$. The charges are $i$ and $i*$ for one red edges, and $j$ and $j^*$ for the blue ones. Furthermore since the "blue" edges are actually bounding a disk the corresponding charge must satisfy $j = 0$. Since the gluing homeomorphism $h$ is the identity of $T^2$ we just take two copies of these rectangles according to Figure \ref{fig5}, thus defining the "basic" Heegaard splitting of $S_{1} \times S_{2}$.

The sum over labelings then gives:
\bea
\sum\limits_{i,j=0}^{N-1} \delta_{i+j-i-j,0} \, \delta_{j,0} = N \, ,
\eea
and the TV invariant of $S_{1} \times S_{2}$ is:
\bea
\Upsilon_N\left(S_{1}\times S_{2}\right) = \frac{1}{N^{1-1}} N = N \, .
\eea

3) The lens space $L(2,1) = {\mathbb{R}}P^{3}$. We present graphically the result knowing that the reasoning is very similar to the previous case. however in this case the decompositions are not the same in the two copies of $T^2$.

\begin{figure}
  \centering
  \includegraphics[scale=0.5]{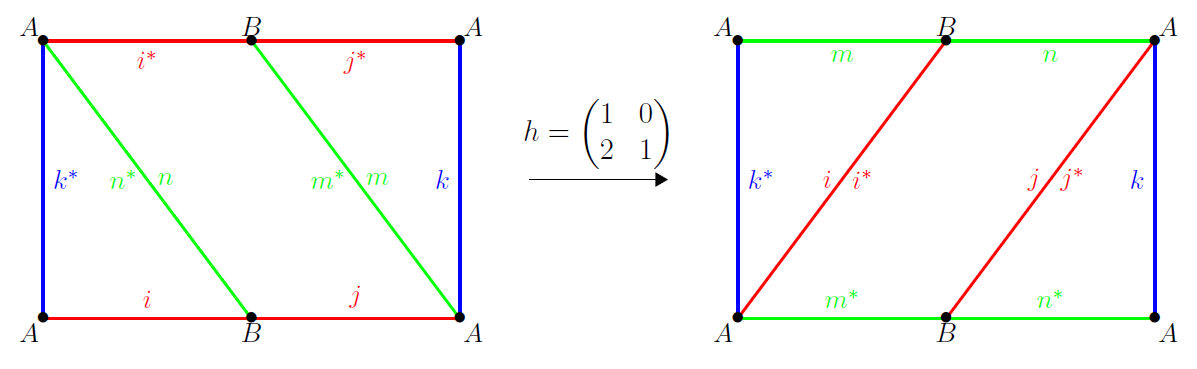}
  \caption{Polyhedral decomposition of $T^2$ compatible with a Heegaard splitting of $L(2,1)$ along $T^2$.}\label{fig6}
\end{figure}

The sum over labelings gives:
\bea
\sum\limits_{i,j,k,m,n=0}^{N-1} \delta_{i+j,0} \, \delta_{m+n,0} \, \delta_{i-k-n,0}  &\,& \hspace{-9.3mm}   \delta_{-j+k+m,0} \, \delta_{-i+j-m+n,0} \\
&=& \sum\limits_{j,k,m=0}^{N-1} \delta_{2\left(j+n\right),0} \, \delta_{j+k+n,0} \, \delta_{j-k-n,0} \\
&=& \sum\limits_{j,m=0}^{N-1} \delta_{2\left(j+m\right),0} \\
&=& N \gcd\left(N,2\right) \, ,
\eea
thus yielding the TV invariant of $L(2,1)$:
\bea
\Upsilon_N\left({\mathbb{R}}P^{3}\right) = \frac{1}{N^{2-1}} N \gcd\left(N,2\right) = \gcd\left(N,2\right) \, .
\eea

\section{Conclusion}

\noindent Let us recall that in the non-abelian (typically $SU(2)$) case (1) the RT invariant and CS partition function coincide; (2) so do the TV invariant and the BF partition function; (3) the absolute square of the RT invariant is also the TV invariant; (4) the absolute square of the CS partition function is the  BF partition function. Point (3) occurs because the category underlying the construction of the RT invariant is modular; point (4) occurs because the space of gauge classes of $SU(2)$-connections on a closed $3$-manifold is connected so that at the level of functional integration we can reparameterize the BF partition function in order to turn it into the absolute square of the CS partition function. As this article shows it in details, of the above points only point (2) holds true in the $U(1)$ case. Firstly the category on which is based the construction of the abelian RT invariant is not modular when this invariant is related to the $U(1)$ CS partition function. Secondly the space of gauge classes of $U(1)$-connections on $M$ is not connected which prevents the $U(1)$ BF partition function from being the absolute square of the CS one. When the category underlying the construction of the RT invariant is modular we recover that the corresponding TV invariant is the absolute square of the RT invariant. There is even a case where there are neither RT invariant nor $U(1)$ CS theory although a TV invariant and a $U(1)$ BF theory exist. As mentioned before, whereas the $U(1)$ case might appear as a trivial one, it actually turns out to be quite tricky. Fortunately Deligne-Belinson cohomology is once more a powerful tool to investigate properly the relation between $U(1)$ CS and BF theories.

Furthermore, although the realization of the BF theory as a lattice gauge theory was already known (at least in the non-abelian case) \cite{FL} the definition of the TV invariant given in this article shows it very clearly. The topological character of the BF theory thus appears through the cohomological nature of TV invariant.

The last step in the study of the $U(1)$ BF theory on closed $3$-dimensional manifolds is to deal with observables. They are given by holonomies of the connections $A$ and $B$ and there expectation values might provide new interesting result compared to the well-known $U(1)$ CS theory \cite{MT}.

The generalization to $4n+3$-dimensional closed manifolds of the results obtained in this article is straightforward. We could also try to adapt the DB approach proposed here in order to study the case of manifolds with boundary \cite{HL}.

As a final remark, let us point out that the $U(1)$ BF theory presented in this article is not the same as the one considered in \cite{Ca}. In this latter article the author considers on a 3-manifold $M$ the lagrangian $B \wedge (dA + \omega \wedge A)$ where $A$, $B$ and $\omega$ are all $1$-forms (closed for $\omega$) on $M$ whereas we know that this lagrangian is ill-defined at $\omega = 0$ if $A$ and $B$ are $U(1)$ connections on a generic closed 3-manifold.

\vskip 1 truecm

{\bf{Acknowledgments}} The authors would like to thank Matthieu Vanicat, Eric Ragoucy and Luc Frappat for fruitful discussions concerning Quantum Groups. We also thank Vladimir Turaev for having kindly answered some of our questions concerning the construction of the TV invariant.

\vfill\eject

\section{Appendix}

\noindent Let's remind a few basics of group theory that we need now. The homology group $H_{1}$ is an Abelian group and Abelian groups are generically classified in such a way that $H_{1}$ can be decomposed as $H_{1} = F_{1} \oplus T_{1}$ with the so-called "free part" $F_{1}=\mathbb{Z}^{b_{1}}$ ($b_{1}$ being the first Betti number) and the so-called "torsion part" $T_{1}= \mathbb{Z}_{p_{1}} \oplus\cdots \oplus \mathbb{Z}_{p_{N}}$ such that $p_{1} \mid p_{2}$, $p_{2}\mid p_{3}$, $\cdots$, $p_{N-1} \mid p_{N}$.

The symmetric bilinear linking form $Q: T_1(M) \rightarrow T_1(M)$ is non-degenerate \cite{Gr}. Its associated matrix $\left(Q_{ij}\right)_{1\leq i,j\leq N}$ is such that $Q_{ij}=\frac{q_{ij}}{p_{i}}$ (see for instance equation (\ref{squaretor})) with $q_{ij} \in \mathbb{Z}$, $\frac{q_{ij}}{p_{i}}=\frac{q_{ji}}{p_{j}}$ and $\gcd\left(q_{ii},p_{i}\right)=1$. This last condition ensures that the torsion cycles chosen to define the matrix elements $Q_{ij}$ are appropriate generators of $T_1(M)$. This is why, for instance, $p$ and $q$ are chosen to be coprime integers in the lens space $L(p,q)$.

For any integer $k$, we denote $p'_{i}$ the integer such that $p_{i}=p'_{i}\gcd\left(k,p_{i}\right)$ and $k'_{i}$ the integer such that $k=k'_{i}\gcd\left(k,p_{i}\right)$. We suppose also that $p'_{i}$ is odd for $1 \leq i \leq \alpha$, pure multiple of $2$ for $\alpha + 1 \leq i \leq \alpha + \beta$ and multiple of $4$ for $\alpha + \beta + 1 \leq i \leq \alpha + \beta + \gamma = N$.

Let's define $\vec{1} = \left(1,\cdots,1\right)$ and $\vec{p} = \left(p_{1},\cdots,p_{N}\right)$ and we consider in the following the usual euclidean scalar product $\langle \cdot,\cdot\rangle$.

We compute first $Z_{BF_{k}}$:
\bea
Z_{BF_{k}}
&=& \sum\limits_{\vec{n}=\vec{0}}^{\vec{p}-\vec{1}}
\sum\limits_{\vec{m}=\vec{0}}^{\vec{p}-\vec{1}}e^{-2i\pi k \langle \vec{m}, Q \vec{n}\rangle} \\
&=& \sum\limits_{\vec{n}=\vec{0}}^{\vec{p}-\vec{1}} \sum\limits_{\vec{m}=\vec{0}}^{\vec{p}-\vec{1}}e^{-2i\pi \langle \vec{m}, Q \left(k\vec{n}\right)\rangle} \\
Z_{BF_{k}}
&=& \sum\limits_{\vec{n}=\vec{0}}^{\vec{p}-\vec{1}}
\prod\limits_{i=1}^{N}
\left(\sum\limits_{m_{i}=0}^{p_{i}-1}\left(e^{-2i\pi \sum\limits_{j=1}^{N}Q_{ij}k n_{j}}\right)^{m_{i}}\right),
\eea
and for a fixed $\vec{n}$, the term:
\bea
\prod\limits_{i=1}^{N}
\left(\sum\limits_{m_{i}=0}^{p_{i}-1}\left(e^{-2i\pi \sum\limits_{j=1}^{N}Q_{ij}k n_{j}}\right)^{m_{i}}\right),
\eea
is not zero if and only if:
\bea
\forall i\in \left[1,N \right],\, \sum\limits_{j=1}^{N}Q_{ij} k n_{j}\in \mathbb{Z}
\eea
that is to say:
\bea
Q\left(k\vec{n}\right)\egzz \vec{0}
\eea
which admits $\vec{0}$ as only solution since $Q$ is non-degenerate. We have then to solve a new set of equations:
\bea
\forall i \in \left[1,N\right],\, k n_{i} = 0 \left[p_{i}\right]
\eea
which is the same as finding all the $\lambda_{j}$ such that $k n_{j} = \lambda_{j} p_{j}$ and $n_{j} \in \left[0,p_{j}-1\right]$. Hence, $n_{j} = \frac{\lambda_{j} p_{j}}{k} = \frac{\lambda_{j} p'_{j}}{k'_{j}}\in \left[0,p_{j}-1\right]$ for $\lambda_{j}=0,k'_{j},\cdots,\left(\gcd\left(k,p_{j}\right)-1\right)k'_{j}$. Indeed, if $\lambda_{j}=\gcd\left(k,p_{j}\right)k'_{j}$ then $n_{j} = p_{j}\notin\left[0,p_{j}-1\right]$. There are thus $\prod\limits_{j=1}^{N}\gcd\left(k,p_{j}\right)$ solutions to the previous set of equations. For each one of those solutions,
\bea
\prod\limits_{i=1}^{N}
\left(\sum\limits_{m_{i}=0}^{p_{i}-1}\left(e^{-2 i \pi \sum\limits_{j=1}^{N}Q_{ij}k n_{j}}\right)^{m_{i}}\right)
=\prod\limits_{i=1}^{N}
\left(\sum\limits_{m_{i}=0}^{p_{i}-1}1\right)
= \prod\limits_{j=1}^{N} p_{j}
\eea

Finally:
\bea
Z_{BF_{k}} = \prod\limits_{j=1}^{N}\gcd\left(k,p_{j}\right)p_{j} \, .
\eea

Concerning $\left| Z_{CS_{k}}(M) \right|^2$, we can write:
\bea
\left| Z_{CS_{k}}(M) \right|^2 = \sum\limits_{\vec{m}=\vec{0}}^{\vec{p}-\vec{1}} \sum\limits_{\vec{n} =\vec{0}}^{\vec{p}-\vec{1}} e^{ 2 \pi i \left(\langle \vec{m},Q \vec{m}\rangle - \langle \vec{n},Q \vec{n}\rangle\right)}.
\eea
By first setting $\vec{n'} = \vec{m} + \vec{n}$:
\bea
\nonumber
\left|Z_{CS_{k}}\right|^{2}
&=& \sum\limits_{\vec{m}=\vec{0}}^{\vec{p}-\vec{1}}
\sum\limits_{\vec{n'}=\vec{m}}^{\vec{p}-\vec{1}+\vec{m}}
e^{-2i\pi k \langle \vec{n'}, Q \left(\vec{n'}-2 \vec{m}\right)\rangle}\\
\nonumber
&=& \sum\limits_{\vec{m}=0}^{\vec{p}-\vec{1}}
\sum\limits_{\vec{n'}=\vec{m}}^{\vec{p}-\vec{1}}
e^{-2i\pi k \langle \vec{n'}, Q \left(\vec{n'}-2\vec{m}\right)\rangle}
+\sum\limits_{\vec{m}=0}^{\vec{p}-\vec{1}} \sum\limits_{\vec{n'}=\vec{p}}^{\vec{p}-\vec{1}+\vec{m}}
e^{-2i\pi k \langle \vec{n'}, Q \left(\vec{n'}-2\vec{m}\right)\rangle}\\
\nonumber
&=& \sum\limits_{\vec{m}=\vec{0}}^{\vec{p}-\vec{1}}\sum\limits_{\vec{n'}=\vec{m}}^{\vec{p}-\vec{1}}
e^{-2i\pi k \langle\vec{n'}, Q \left(\vec{n'}-2\vec{m}\right)\rangle}
+\sum\limits_{\vec{m}=\vec{0}}^{\vec{p}-\vec{1}}
\sum\limits_{\vec{n''}=\vec{0}}^{\vec{m}-\vec{1}}
e^{-2i\pi k \langle\left(\vec{n''}+\vec{p}\right), Q \left(\vec{n''}+\vec{p}-2\vec{m}\right)\rangle}\\
\label{period}
&=& \sum\limits_{\vec{m}=\vec{0}}^{\vec{p}-\vec{1}}
\sum\limits_{\vec{n'}=\vec{m}}^{\vec{p}-\vec{1}}
e^{-2i\pi k \langle \vec{n'} , Q \left(\vec{n'}-2\vec{m}\right)\rangle}
+\sum\limits_{\vec{m}=\vec{0}}^{\vec{p}-\vec{1}}
\sum\limits_{\vec{n''}=\vec{0}}^{\vec{m}-\vec{1}}
e^{-2i\pi k \langle \vec{n''} , Q \left(\vec{n''}-2\vec{m}\right)\rangle}\\
\nonumber
&=& \sum\limits_{\vec{m}=\vec{0}}^{\vec{p}-\vec{1}}
\sum\limits_{\vec{n'}=\vec{0}}^{\vec{p}-\vec{1}}
e^{-2i\pi k \langle \vec{n'} , Q \left(\vec{n'}-2\vec{m}\right) \rangle}\\
\nonumber
&=&\sum\limits_{\vec{n'}=\vec{0}}^{\vec{p}-\vec{1}}e^{-2i\pi k \langle\vec{n'}, Q \vec{n'}\rangle}
\sum\limits_{\vec{m}=\vec{0}}^{\vec{p}-\vec{1}}e^{2i\pi k \langle \vec{n'}, Q \left(2\vec{m}\right)\rangle}\\
\label{linsym}
&=&\sum\limits_{\vec{n'}=\vec{0}}^{\vec{p}-\vec{1}}e^{-2i\pi k \langle\vec{n'}, Q \vec{n'}\rangle}
\sum\limits_{\vec{m}=\vec{0}}^{\vec{p}-\vec{1}}e^{2i\pi k \langle \vec{m}, Q \left(2\vec{n'}\right)\rangle}\\
\nonumber
\left|Z_{CS_{k}}\right|^{2}
&=&\sum\limits_{\vec{n'}=\vec{0}}^{\vec{p}-\vec{1}}e^{-2i\pi k \langle\vec{n'}, Q \vec{n'}\rangle}
\prod\limits_{i=1}^{N}
\left(\sum\limits_{m_{i}=0}^{p_{i}-1}\left(e^{2i\pi k Q \left(2\vec{n'}\right)}\right)^{m_{i}}\right),
\eea
where we used the periodicity of $Q$ for (\ref{period}) then its symmetry and linearity for (\ref{linsym}).

For a fixed $\vec{n}$, the term:
\bea
\prod\limits_{i=1}^{N}
\left(\sum\limits_{m_{i}=0}^{p_{i}-1}\left(e^{-2i\pi k Q\left( 2 \vec{n}\right)}\right)^{m_{i}}\right)
= \prod\limits_{i=1}^{N}
\left(\sum\limits_{m_{i}=0}^{p_{i}-1}\left(e^{-2i\pi \sum\limits_{j=1}^{N}Q_{ij}2k n_{j}}\right)^{m_{i}}\right),
\eea
is not zero if and only if:
\bea
\forall i\in \left[1,N \right],\, \sum\limits_{j=1}^{N}Q_{ij} 2k n_{j}\in \mathbb{Z}
\eea
that is to say:
\bea
Q\left(2k\vec{n}\right)\egzz \vec{0}
\eea
which admits $\vec{0}$ as only solution since $Q$ is non-degenerate. We have then to solve a new set of equations:
\bea
\forall i \in \left[1,N\right],\, 2k n_{i} = 0 \left[p_{i}\right]
\eea
which is the same as finding all the $\lambda_{j}$ such that $2k n_{j} = \lambda_{j} p_{j}$ and $n_{j} \in \left[0,p_{j}-1\right]$. Hence, $n_{j} = \frac{\lambda_{j} p_{j}}{2k} = \frac{\lambda_{j} p'_{j}}{2k'_{j}}\in \left[0,p_{j}-1\right]$. Starting from here, we have to distinguish the case where $1 \leq j \leq \alpha$ and $\alpha+1 \leq j \leq N$ that is to say the case where $p'_{j}$ is odd and $p'_{j}$ is even.

First for $1 \leq j \leq \alpha$, $p'_{j}$ is odd, so $2k'_{j}\mid\lambda_{j}$ and as a consequence, $\lambda_{j} = 0, 2k'_{j}, 4k'_{j}, \cdots,$ $\left(\gcd\left(k,p_{j}\right)-1\right)2k'_{j}$, which gives a total of $\gcd\left(k,p_{j}\right)$ solutions that we will now write $n_{j} = \mu_{j}p'_{j}$ with $\mu_{j}\in\left[0,\gcd\left(k,p_{j}\right)-1\right]$.

Secondly for $\alpha+1 \leq j \leq N$, $p'_{j}$ is even (note that this implies that $k'_{j}$ is odd), so $n_{j} = \frac{\lambda_{j} p'_{j}}{2k'_{j}} = \frac{\lambda_{j} p''_{j}}{k'_{j}}$ and as a consequence, $\lambda_{j} = 0, k'_{j}, 2k'_{j}, \cdots, \left(2\gcd\left(k,p_{j}\right)-1\right)k'_{j}$, which gives a total of $2\gcd\left(k,p_{j}\right)$ solutions that we will now write $n_{j} = \nu_{j}p''_{j}$ with $\nu_{j}\in\left[0,\gcd\left(k,p_{j}\right)-1\right]$.

We thus determined a set $S$ of solutions $\vec{n} = \left(n_{1},\cdots,n_{N}\right)$ whose cardinal is:
\bea
2^{\beta+\gamma}\prod\limits_{j=1}^{\alpha+\beta+\gamma=N}\gcd\left(k,p_{j}\right).
\eea

Hence we can write:
\bea
\left|Z_{CS_{k}}\right|^{2}
=\prod\limits_{j=1}^{N}p_{j}\sum\limits_{\vec{n} \in S}e^{2i\pi k\langle\vec{n},Q\vec{n}\rangle}
=\prod\limits_{j=1}^{N}p_{j}\sum\limits_{\vec{n} \in S}e^{2i\pi \langle\vec{n},Q\left(k\vec{n}\right)\rangle}.
\eea
Let's consider:
\bea
\langle\vec{n},Q\left(k\vec{n}\right)\rangle
=\sum\limits_{i=1}^{N}k n_{i}^{2}Q_{ii}
+2\sum\limits_{1\leq i<j\leq N}^{N}k n_{i} n_{j}Q_{ij}
\eea
that we can rewrite:
\bea
\langle\vec{n},Q\left(k\vec{n}\right)\rangle
&=&\sum\limits_{i=1}^{N}k \xi_{i}^{2}\epsilon_{i}^{2}p'^{2}_{i}\frac{q_{ii}}{p_{i}}
+2\sum\limits_{1\leq i<j\leq N}^{N}k \xi_{i}\xi_{j}\epsilon_{i}\epsilon_{j}p'_{i}p'_{j}\frac{q_{ij}}{p_{i}}\\
&=&\sum\limits_{i=1}^{N}k'_{i} \xi_{i}^{2}\epsilon_{i}^{2}p'_{i}q_{ii}
+2\sum\limits_{1\leq i<j\leq N}^{N}k'_{i} \xi_{i}\xi_{j}\epsilon_{i}\epsilon_{j}p'_{j}q_{ij}
\eea
with $\xi_{i} = \mu_{i}$ if $1\leq i \leq \alpha$ and $\xi_{i} = \nu_{i}$ if $\alpha+1 \leq i \leq \alpha + \beta + \gamma=N$ and $\epsilon_{i} = 1$ if $1\leq i \leq \alpha$ and $\epsilon_{i} = \slfrac{1}{2}$ if $\alpha+1 \leq i \leq \alpha + \beta + \gamma=N$.

We start with the second term, corresponding to the non-diagonal terms of $Q$:
\bea
2\sum\limits_{1\leq i<j\leq N}^{N}k'_{i}\xi_{i}\xi_{j}\epsilon_{i}\epsilon_{j}p'_{j}q_{ij} \egzz 0
\eea
since if there is $\epsilon_{j}=\slfrac{1}{2}$ then $p'_{j}$ is even and thus cancels $\epsilon_{j}$ and if $\epsilon_{i}=\slfrac{1}{2}$ also, then it is cancelled by the factor $2$ in front of the sum. Hence, we see that the diagonal terms cannot contribute in the exponential.

Concerning the first term, corresponding to the diagonal terms of $Q$:
\bea
\label{diago}
\sum\limits_{i=1}^{N}k'_{i} \xi_{i}^{2}\epsilon_{i}^{2}p'_{i}q_{ii}
\nonumber
&=&\sum\limits_{i=1}^{\alpha}k'_{i} \xi_{i}^{2}\epsilon_{i}^{2}p'_{i}q_{ii}
+\sum\limits_{i=\alpha+1}^{\alpha+\beta}k'_{i} \xi_{i}^{2}\epsilon_{i}^{2}p'_{i}q_{ii}
+\sum\limits_{i=\alpha+\beta+1}^{\alpha+\beta+\gamma=N}k'_{i} \xi_{i}^{2}\epsilon_{i}^{2}p'_{i}q_{ii} \\
\nonumber
&\egzz& \sum\limits_{i=\alpha+1}^{\alpha+\beta}k'_{i} \xi_{i}^{2}\epsilon_{i}^{2}p'_{i}q_{ii} = \sum\limits_{i=\alpha+1}^{\alpha+\beta}k'_{i}\epsilon_{i}\xi_{i}^{2}p''_{i}q_{ii} \\
&\egzz& \frac{1}{2}\sum\limits_{i=\alpha+1}^{\alpha+\beta}k'_{i}\xi_{i}^{2}p''_{i}q_{ii}
\eea
since the first term in the right hand side of (\ref{diago}) is trivially an integer and the third one too in so far as $p'_{i}$ is multiple of $4$ for $i \in \left[\alpha+\beta+1,\alpha+\beta+\gamma\right]$.

We have now to consider the case $\beta = 0$ and the case $\beta \neq 0$. The first one is simple, since $\langle\vec{n},Q\left(k\vec{n}\right)\rangle \egzz 0$ and as a result:
\bea
\left|Z_{CS_{k}}\right|^{2}
=\prod\limits_{j=1}^{N}p_{j}\sum\limits_{\vec{n} \in S}1 = \left|S\right| = 2^{\gamma}\prod\limits_{j=1}^{N}p_{j}\gcd\left(k,p_{j}\right).
\eea

Moreover, if $\beta \neq 0$, for $i \in \left[\alpha,\alpha+\beta+1\right]$, necessarily $k'_{i}$, $p''_{i}$ and $q_{ii}$ are odd, thus their product too, and as a consequence:
\bea
\label{5.90}
\frac{1}{2}\sum\limits_{i=\alpha+1}^{\alpha+\beta}k'_{i}\xi_{i}^{2}p''_{i}q_{ii}
=\frac{1}{2}\sum\limits_{i=\alpha+1}^{\alpha+\beta}\xi_{i}^{2}\left(2\rho_{i}+1\right)
\egzz\frac{1}{2}\sum\limits_{i=\alpha+1}^{\alpha+\beta}\xi_{i}^{2}
\eea
and hence we are interested in the parity of the sum:
\bea
\sum\limits_{i=\alpha+1}^{\alpha+\beta}\xi_{i}^{2}
\eea
This sum is odd if and only if it contains an odd number of odd $\xi_{i}$. For that we have:
\bea
\sum\limits_{k=0 \atop 2\nmid k}^{\beta} {\beta \choose k}=2^{\beta-1}
\eea
possibilities in the choice of the indices, then $2^{\gamma}\prod\limits_{j=1}^{N}\gcd\left(k,p_{j}\right)$ possibilites in the choice of the numbers of those corresponding indices. This gives a subset $S_1$ of $S$ of cardinal $2^{\beta-1}2^{\gamma}\prod\limits_{j=1}^{N}\gcd\left(k,p_{j}\right) = \frac{\left|S\right|}{2}$. Let's call $S_{2}$ the complement of $S_{1}$ in $S$, which thus contains all the $\vec{n}$ of $S$ which give an even sum. We remark that $\left|S_{2}\right| = \left|S_{1}\right| = \frac{\left|S\right|}{2}$.

Finally we can write:
\bea
\left|Z_{CS_{k}}\right|^{2}
\nonumber
&=&\prod\limits_{j=1}^{N}p_{j}\sum\limits_{\vec{n} \in S}e^{2i\pi \langle\vec{n},Q\left(k\vec{n}\right)\rangle}\\
\nonumber
&=&\prod\limits_{j=1}^{N}p_{j}\sum\limits_{\vec{n} \in S}e^{2i\pi \langle\vec{n},Q\left(k\vec{n}\right)\rangle}\\
\nonumber
&=&\prod\limits_{j=1}^{N}p_{j}\left(\sum\limits_{\vec{n} \in S_{1}}e^{2i\pi \langle\vec{n},Q\left(k\vec{n}\right)\rangle}+\sum\limits_{\vec{n} \in S_{2}}e^{2i\pi \langle\vec{n},Q\left(k\vec{n}\right)\rangle}\right)\\
\nonumber
&=&\prod\limits_{j=1}^{N}p_{j}\left(\sum\limits_{\vec{n} \in S_{1}}\left(-1\right)+\sum\limits_{\vec{n} \in S_{2}}1\right)\\
\nonumber
&=&\prod\limits_{j=1}^{N}p_{j}\left(-\left|S_{1}\right|+\left|S_{2}\right|\right)\\
\left|Z_{CS_{k}}\right|^{2}
&=& 0.
\eea
Let us point out that the sum (\ref{5.90}) is $1/2$ (modulo integers) exactly when there is a $\alpha \in H^1(M,\mathbb{Z}_{2k})$ such that $\alpha \Cupp \alpha \Cupp \alpha \neq 0$. Indeed since $\frac{1}{2k}(\alpha \Cupp \alpha \Cupp \alpha)(M) = k Q(\bar{\alpha},\overline{\alpha})$ where $\bar{\alpha} \in T_1$ is defined by $Q(\bar{\alpha},\bar{u}) = \frac{1}{2k} \alpha(\bar{u})$ for all $\bar{u} \in T_1$ \cite{Tu2,MOO}, we see that $(\alpha \Cupp \alpha \Cupp \alpha)(M) = 0$ in $\mathbb{Z}_{2k}$ if and only if $k Q(\bar{\alpha},\overline{\alpha}) \in \mathbb{Z}$ which holds if and only if the sum (\ref{5.90}) is an integer. Finally this means that $\beta = 0$ in our computation.

\vfill\eject

\end{document}